\documentclass[prc,showpacs,groupedaddress,twocolumn]{revtex4-1}

\PassOptionsToPackage{merge}{natbib}
\usepackage{xspace}
\usepackage{amsmath}
\usepackage{graphicx}
\usepackage{graphics}
\usepackage{hyperref}

\begin{document}

\newcommand{\tcu}{$^{63}$Cu\xspace}
\newcommand{\fcu}{$^{65}$Cu\xspace}
\newcommand{\majorana}{M{\textsc{ajorana}} \xspace}
\newcommand{\grays}{$\gamma$-rays \xspace}
\newcommand{\gray}{$\gamma$-ray \xspace}
\newcommand{\nnprime}{(n,n$^\prime \gamma$)\xspace}
\newcommand{\ntwon}{n,2n$\gamma$\xspace}
\newcommand{\nthreen}{n,3n$\gamma$\xspace}
\newcommand{\nxn}{n,xn$\gamma$\xspace}
\newcommand{\nx}{n,x$\gamma$\xspace}

\newcommand{\natpb}{$^{\textrm{nat}}$Pb\xspace}
\newcommand{\natcu}{$^{\textrm{nat}}$Cu\xspace}
\newcommand{\natge}{$^{\textrm{nat}}$Ge\xspace}
\newcommand{\nonubb}  {$0 \nu \beta \beta$\xspace}
\newcommand{\twonubb} {$2 \nu \beta \beta$\xspace}
\newcommand{\gam}{$\gamma$}
\def\nuc#1#2{${}^{#1}$#2}
\def\mee{$\langle m_{\beta\beta} \rangle$}
\def\mnu{$\langle m_{\nu} \rangle$}
\def\ml{$m_{lightest}$}
\def\gnu{$\langle g_{\nu,\chi}\rangle$}
\def\mmod{$\| \langle m_{\beta\beta} \rangle \|$}
\def\mb{$\langle m_{\beta} \rangle$}
\def\BBz{$0 \nu \beta \beta$}
\def\BBm{$\beta\beta(0\nu,\chi)$}
\def\BBt{$2 \nu \beta \beta$}
\def\nonubb{$0 \nu \beta \beta$}
\def\twonubb{$2 \nu \beta \beta$}
\def\BB{$\beta\beta$}
\def\Mz{$M_{0\nu}$}
\def\Mt{$M_{2\nu}$}
\def\MzG{$M^{GT}_{0\nu}$}           
\def\MzF{$M^{F}_{0\nu}$}                
\def\MtG{$M^{GT}_{2\nu}$}           
\def\MtF{$M^{F}_{2\nu}$}                
\def\Tz{$T^{0\nu}_{1/2}$}
\def\Tt{$T^{2\nu}_{1/2}$}
\def\Tc{$T^{0\nu\,\chi}_{1/2}$}
\def\Rz{$\Gamma_{0\nu}$}            
\def\Rt{$\Gamma_{2\nu}$}            
\def\ms{$\delta m_{\rm sol}^{2}$}
\def\ma{$\delta m_{\rm atm}^{2}$}
\def\ts{$\theta_{\rm sol}$}
\def\ta{$\theta_{\rm atm}$}
\def\tot{$\theta_{13}$}
\def\gpp{$g_{pp}$}                  
\def\qval{$Q_{\beta\beta}$}                 
\def\MJ{{\sc Majorana}}             
\def\DEM{{\sc Demonstrator}}             
\def\be{\begin{equation}}
\def\ee{\end{equation}}
\def\cpRty{counts/ROI/t-y}
\def\onecpRty{1~count/ROI/t-y}
\def\fourcpRty{4~counts/ROI/t-y}
\def\ppc{P-PC}                          
\def\nsc{N-SC}                          

\title{Neutron inelastic scattering in natural Cu as a background in neutrinoless double-beta decay experiments}

\newcommand{\lanl}{Physics Division, Los Alamos National Laboratory, Los Alamos, NM 87545}
\newcommand{\lansce}{LANSCE Division, Los Alamos National Laboratory, Los Alamos, NM 87545}
\newcommand{\usd}{Physics Department, University of South Dakota, Vermillion, South Dakota 57069}
\newcommand{\Columbia}{Present Address: Department of Physics, Columbia University, New York, NY 10027}
\newcommand{\tlanl}{Theory Division, Los Alamos National Laboratory, Los Alamos, NM 87545}

\affiliation{\lanl}
\affiliation{\usd}
\affiliation{\lansce}
\affiliation{\Columbia}

\author{M.S.~Boswell}\email[]{mitzib@lanl.gov}\affiliation{\lanl}
\author{M. Devlin}\affiliation{\lansce}
\author{S.R.~Elliott}\affiliation{\lanl}
\author{T. Kawano}\affiliation{\tlanl}
\author{N. Fotiades}\affiliation{\lansce}
\author{V.E.~Guiseppe}\affiliation{\usd}
\author{R.O.~Nelson}\affiliation{\lansce}
\author{D.V.~Perepelitsa}\altaffiliation{\Columbia}\affiliation{\lanl}
\date{\today}

\begin{abstract}
{\bf{Background:}}Experiments designed to study rare processes, such as neutrinoless double beta decay ($0\nu\beta\beta$), are crucial tests for physics beyond the standard model.   These experiments rely on reducing the intrinsic radioactive background to unprecedented levels, while adequately shielding the detectors from external sources of radioactivity. {\bf{Purpose:}} An understanding of the potential for neutron excitation of the shielding and detector materials is important for obtaining this level of sensitivity.  {\bf{Methods:}} Using the broad-spectrum neutron beam at LANSCE, we have measured inelastic neutron scattering on $^{nat}$Cu.  The goal of this work is focused on understanding the background rates from neutrons interacting in these materials in regions around the Q-values of many candidate $0\nu\beta\beta$ decay isotopes, as well as providing data for benchmarking Monte Carlo simulations of background events.   {\bf{Results:}} We extracted the level cross sections from the $\gamma$ production cross section for 46 energy levels in $^{nat}$Cu . These level cross sections were compared with the available experimental data, as well as the ENDF/B-VII evaluation for discrete levels. {\bf{Conclusions:}} For energy levels above 2 MeV we found significant discrepancies between the suggested level cross sections for both nuclei and our data.  We found reasonable agreement between our measurement and the ENDF/B-VII evaluation for the total neutron inelastic cross section in \tcu.  Our measurement of the total neutron inelastic scattering cross section in \fcu was 30\% lower than the ENDF/B-VII evaluations, which we attribute to unobserved transitions in \fcu.  Furthermore, we found that the implementation of the ENDF/B-VII evaluation in simulations did not properly model the decay properties of the nucleus to the degree necessary for estimating backgrounds in rear-event searches.  Finally, we examined the potential implications of our measurements on $0\nu\beta\beta$ measurements and found that many of the commonly studied $0\nu\beta\beta$ isotopes had Q-values below the cutoff for ENDF/B-VII evaluated discrete levels in either Cu nucleus. 
\end{abstract}


\maketitle

\section{Introduction}

Neutrinoless double-beta decay (\BBz) plays a key role in understanding the neutrino's absolute mass scale and particle-antiparticle nature~\cite{Ell02,Ell04, Bar04, Avi05, eji05, avi08}. If this nuclear decay process exists, one would observe a mono-energetic line originating from a material containing an isotope subject to this decay mode. One such isotope that may undergo this decay is $^{76}$Ge.
An experiment using germanium-diode detectors fabricated from material enriched in $^{76}$Ge  was the first experiment to established a half-life limit and a restrictive constraints on the effective Majorana mass for the neutrino~\cite{aal02a,bau99}. One analysis~\cite{kla06} of the data in Ref.~\cite{bau99} claims evidence for the decay with a half-life of $2.23^{+0.44}_{-0.31} \times 10^{25}$ y.
Planned Ge-based \BBz\ experiments~\cite{ell08,Gerda} will test this claim. Eventually, these future experiments target a sensitivity of $>$10$^{27}$ y or $\sim$1 event/ton-year to explore neutrino mass values near that indicated by the atmospheric neutrino oscillation results\cite{fog11}. 

The key to these experiments lies in the ability to reduce intrinsic radioactive
background to unprecedented levels and to adequately shield the detectors from external
 sources of radioactivity. Previous experiments' limiting backgrounds have been trace levels of natural decay chain isotopes within the detector and shielding components. The \gam-ray emissions from these isotopes can deposit energy in the Ge detectors producing a continuum, which may overwhelm the potential \BBz\ signal peak at 2039 keV. Great progress has been made identifying the location and origin of this contamination, and future efforts will substantially reduce this contribution to the background. The background level goal of 1 event/ton-year, however, is an ambitious factor better than the currently best achieved background level for Ge-based experiments~\cite{bau99}. If the efforts to reduce the natural decay chain isotopes are successful, previously unimportant components of the background must be understood and eliminated. The work of Mei and Hime\cite{mei06} recognized that $(n,n'\gamma)$ reactions will become important for ton-scale double-beta decay experiments.
 
 Reference~\cite{mei08} recognized that the specific \gam\ rays from Pb isotopes at 2041 and 3062 keV are particularly troublesome. The former is dangerously near the 2039.00$\pm$0.05-keV Q-value for zero-neutrino double-beta decay in $^{76}$Ge and the latter can produce a double-escape peak line at 2040 keV. That paper pointed out that the cross sections to produce these lines in \natpb\ were unmeasured and hence set to zero in the data bases of the simulation codes used to design and analyze \nonubb\ data. This result indicated the importance of assessing all materials used in \nonubb\ experiments for such problematic lines. In this work we examine Cu in this context, because it is a material frequently used in large quantities in \nonubb\ experiments.  
 
 Previous authors have studied \nnprime\ reactions in Cu usually using both natural \citep{bon59,bey56,1968DA,1970Fe,1971Fr}\cite{kram89,zhou,almen,nishimura,tucker,joensson,xiamin,guenther} and isotopic samples \cite{1982De,1983Di,1987Do,1982el,kos01,kosyak00} to isolate the isotopic effects. Previous measurements, however, have not extended to nuclear excitation energies greater than 1.5 MeV and therefore the production rates of $\gamma$ rays beyond this energy were not measured.   
This paper presents measurements of \natcu\nnprime\  production cross sections of the \gam\ rays  for \natcu.  Although our work was motivated by neutron reaction considerations in materials that play important roles in the \MJ\  \cite{avi07} design, the results have wider utility since Cu is used by numerous low-background experiments.

\section{Experimental Procedure and Data Analysis}
The experimental data were collected at the Los Alamos Neutron Science Center (LANSCE) located at Los Alamos National Laboratory.  Spallation neutrons are produced from an 800-MeV proton beam interacting with a $^{nat}$W spallation target in the Weapons Neutron Research (WNR) Facility \cite{1990Lis}.   The incident proton beam is delivered in short pulses spaced 1.8 $\mu$s for 625 $\mu$s at a rate of 40 Hz. The resulting neutrons range in energy from 0.2-800 MeV.  To ensure that the beam area illuminates only the Cu target, the beam is collimated to a 1.9-cm radius with a lead collimator.  The neutron flux is monitored using an in-beam fission chamber with $^{235,238}$U foils.  

	The $^{nat}$Cu target was placed at the center of GEANIE spectrometer \cite{2005Be}, 20.34-m downstream from the production spallation target, and 60$^o$ to the right of the proton beam.  Three stacked foils of $^{nat}$Cu were placed perpendicular to the neutron beam.  The foils, each measuring 50 x 50 x 0.5 mm, had a total mass of 139 g.  The surrounding GEANIE spectrometer comprises 26 germanium detectors;  16 coaxial detectors with an energy range of four MeV, and 10 Compton suppressed planar detectors with an energy range of one MeV.  During the run cycle, the detectors experience significant neutron exposure, and subsequently some of the detectors have reduced energy resolution, or gain instabilities.  During the course of the experiment three planar detectors and six coaxial detectors (nine total detectors) were not used for exhibiting such problematic behavior.   The remaining six planar and ten coaxial detectors were used for this analysis.
	
	The procedure for extracting \gray production cross sections from the GEANIE spectrometer has been discussed in great detail in several papers (see for example  \cite{2009Gui,fot04}.).  The GEANIE spectrometer provides the yields and energies of discrete cascade \gam\ rays.  Yields are converted into partial $\gamma$-production cross-sections using calibrated detector efficiency, deadtime corrections, neutron flux, and target thickness.

\section{Results and Discussion}

	The $\gamma$ production cross sections were the primary measured quantities of the present experiment.   In total 119 \gray transitions were measured in $^{63,64,65}$Cu for neutron energies from 0.2-100 MeV.   The total inelastic and the level cross sections for individual excited states were deduced from the evaluated level scheme of \tcu, $^{64}$Cu, and \fcu.  The level cross sections for selected levels in both Cu nuclei are presented in the sections below, as well as the total neutron inelastic cross sections and (n,2n$\gamma$) cross sections in Cu.    Since the goal of this work is to place these measurements in the context of the typical background simulation for a neutrinoless double beta decay experiment,  we begin our analysis by presenting our data in the context of data from the literature, and the evaluated cross sections presented in the ENDF/B-VII evaluation on Cu.

	The ENDF/B-VII presents cross sections for the excitation of an energy level for a range of incident neutrons.  At high incident neutron energies discrete gamma rays may be produced by various excitation paths, particularly feeding from $\gamma$ decay of higher lying levels.  The neutron inelastic partial cross sections to discrete levels supplied by ENDF/B-VII do not include this important component.  These level excitation cross sections differ from those obtained through individual $\gamma$ production cross sections for a given excited state because the $\gamma$ production cross sections include the effects of feeding from higher energy levels.  Thus level cross sections deduced from summing up the individual $\gamma$ production cross sections associated with a given level are a measure of the level de-excitation cross section.  We have chosen to present these two types of level cross sections together because the focus of this paper is  the accurate simulation of both the excitation and the de-excitation of the energy levels in Cu.  	

	The second section of the results discusses the results of this experiment in the context of two common statistical model calculations, TALYS\cite{talys} \& CoH3\cite{kawano}.   Statistical model calculations are quite important to the data evaluation process, as they serve to fill in the gaps of experimental data and provide cross sections where no measurements are available.   Thereafter we compare the natural Cu\nnprime and (n,2n$\gamma$) cross sections with GEANT4 simulations.  Finally we discuss the implication of this measurement on neutrinoless double beta decay measurements.  
	
\subsection{Comparison with data and the ENDF/B-VII evaluations}

\begin{figure}[tbp]
\begin{center}
\includegraphics[width=0.8\linewidth]{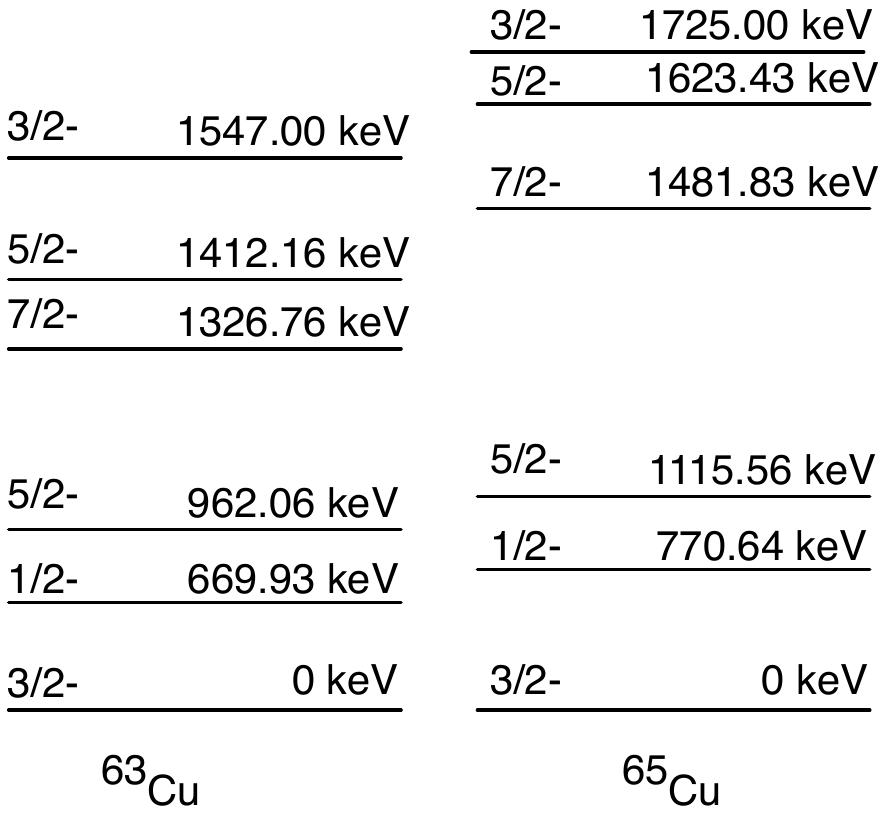}
\caption{The level scheme of the six lowest energy levels in \tcu\cite{erjun2001} and \fcu\cite{browne2010}.  }
\label{fig:level}
\end{center}
\end{figure}

\begin{figure}[htbp]
\begin{center}
\includegraphics[width=\linewidth]{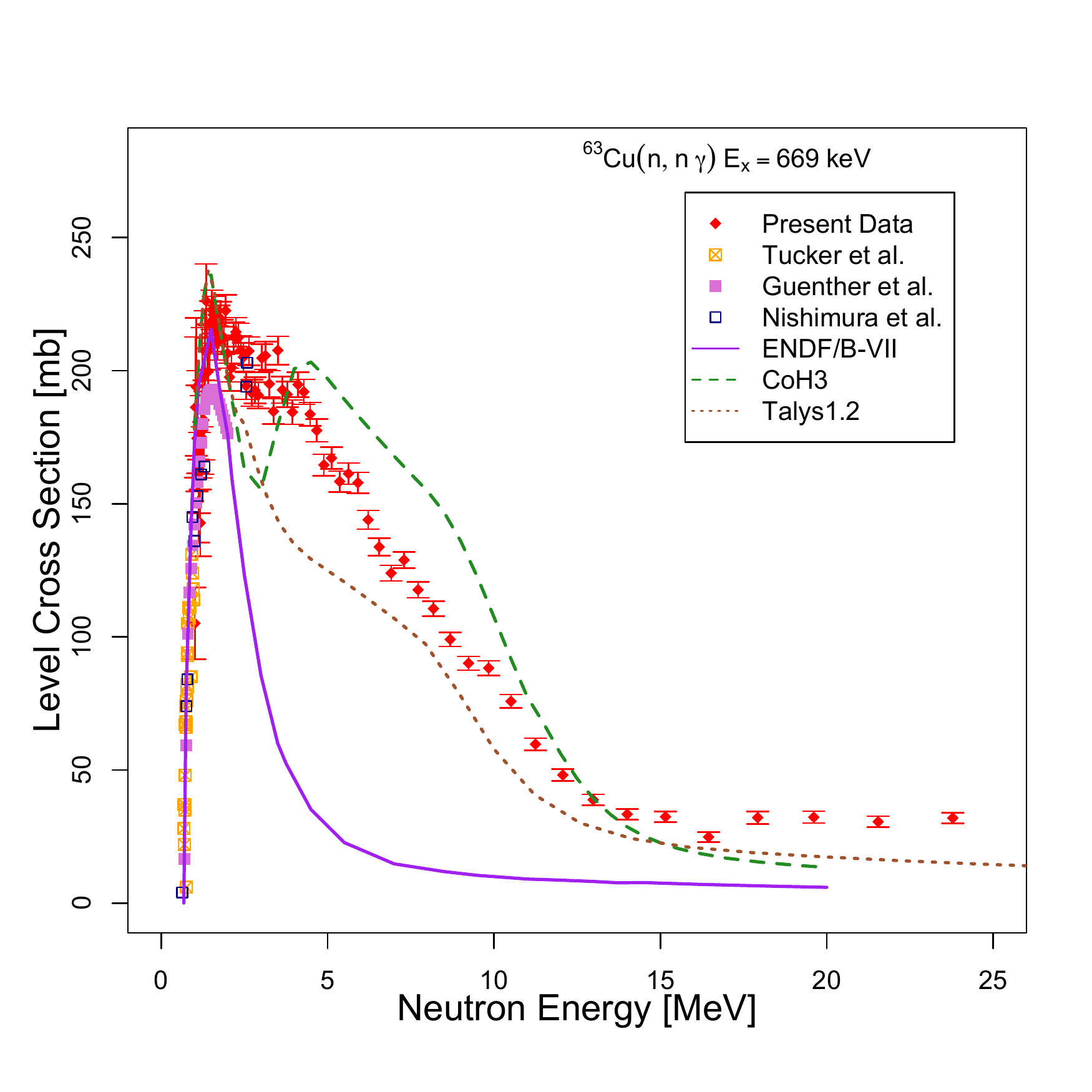}
\caption{The total $\gamma$-ray production cross section for the first excited level (E$_x$=669 keV) in \tcu from this measurement, and several (n,n'$\gamma$) measurements (squares) \cite{guenther,tucker,kram89,nishimura}, the ENDF/B-VII evaluation (solid line),  the TALYS1.2 and CoH3 calculations (dashed lines) for this level.  At high incident neutron energies discrete $\gamma$ rays may be produced by various excitation paths, especially feeding from $\gamma$ decay of higher lying levels, and the neutron inelastic partial cross sections to discrete levels.  Note that the ENDF/B-VII  level cross sections do not include feeding from higher lying levels that is in the measured cross sections and nuclear model calculations.  The level de-excitation cross sections presented for the statistical models are similar to those presented from this measurement; they are calculated by summing up the $\gamma$ production cross section for a particular level.}
\label{fig:ana:63cu669}
\end{center}
\end{figure}

\begin{figure}[htbp]
\begin{center}
\includegraphics[width=\linewidth]{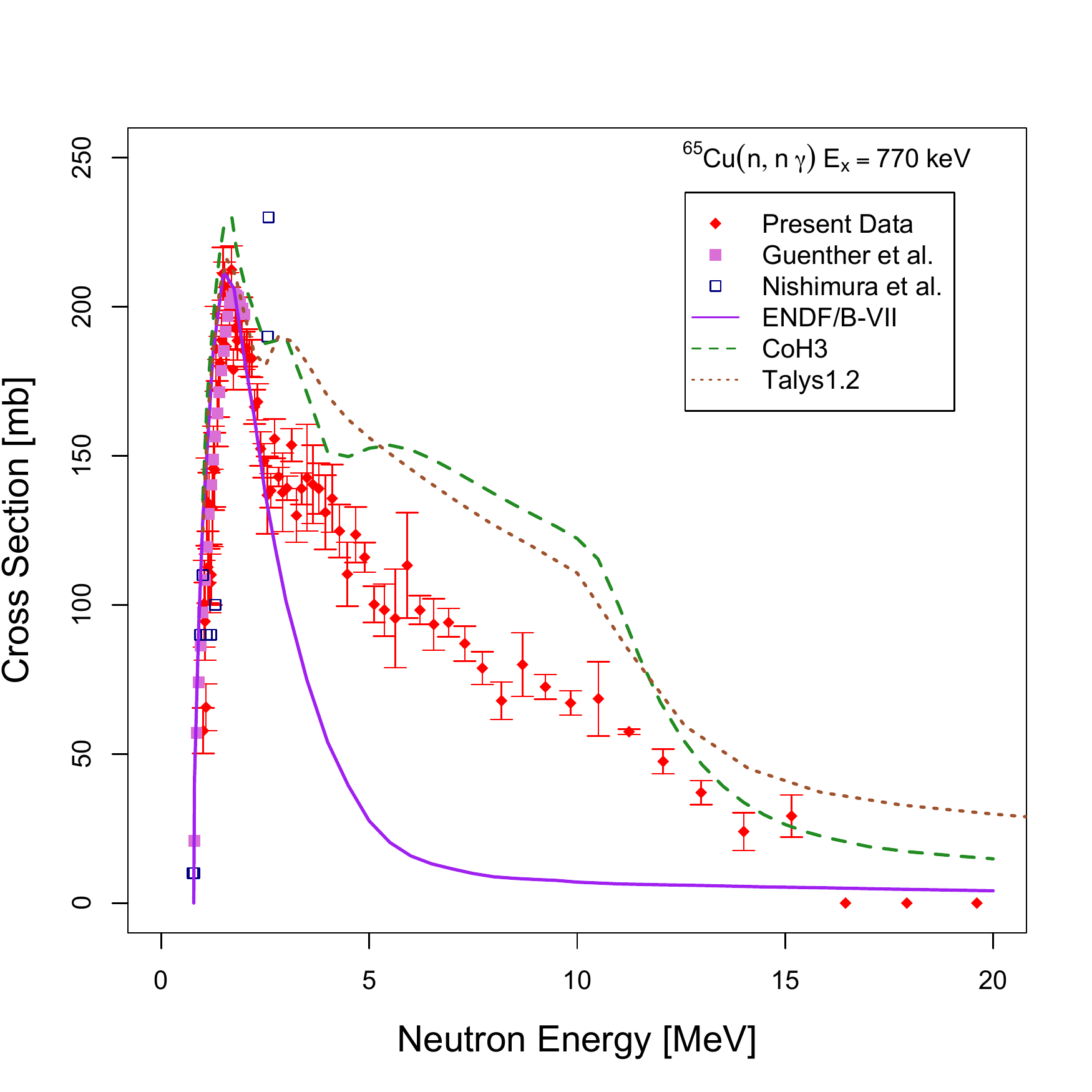}
\caption{The total $\gamma$-ray production cross section  for the first excited level (E$_x$=770 keV) in \fcu from this measurement,  as well as several (n,n'$\gamma$) measurements (squares) \cite{guenther,nishimura}, the ENDF/B-VII evaluation (solid line), the TALYS1.2 and CoH3 calculations (dashed lines) for this level \cite{endf}. For a description of the types of level de-excitation cross sections presented here please see Figure \ref{fig:ana:63cu669}.}
\label{fig:ana:65cu770}
\end{center}
\end{figure}

We extracted the level de-excitation cross sections from the $\gamma$ production cross section for 46 energy levels in $^{nat}$Cu . These level de-excitation cross sections were compared with the available experimental data, as well as the ENDF/B-VII evaluation for level excitation cross section of discrete levels. Figures \ref{fig:ana:63cu669}, and \ref{fig:ana:65cu770} show the level de-excitation cross sections for the first energy levels in \tcu and \fcu.    In these Figs. we show our data along side the ENDF/B-VII evaluations for these levels, as well as several (n,n'$\gamma$) measurements (circles) \cite{guenther,tucker,kram89,nishimura}.  The second energy levels in both nuclei have very weak decay branches to the first excited level, therefore the strength from these indirect excitations doesn't begin to influence the level de-excitation cross section until the neutron energy exceeds the energy required to excite the nucleus into the third excited level (E$_x$=1326 keV for \tcu and 1481 keV for \fcu, see Fig. \ref{fig:level} for the level diagram of both Cu nuclei).  The level de-excitation cross sections peak near 1.34, and 1.689 MeV for \tcu and \fcu, respectively.    For neutron energies below the maximum cross section there is minimal contribution from these higher energy levels, and therefore the level de-excitation cross sections can be entirely attributed to direct excitation.     The measured maximum level de-excitation cross sections are consistent with the ENDF/B-VII evaluation.     The maximum level de-excitation cross section for \tcu is $\sigma_{max}$=226(14) mb, which is roughly 6\% higher than $\sigma_{max}$=212(8) mb for \fcu.  

\begin{figure*}[htbp]
\begin{center}
\includegraphics[width=\linewidth]{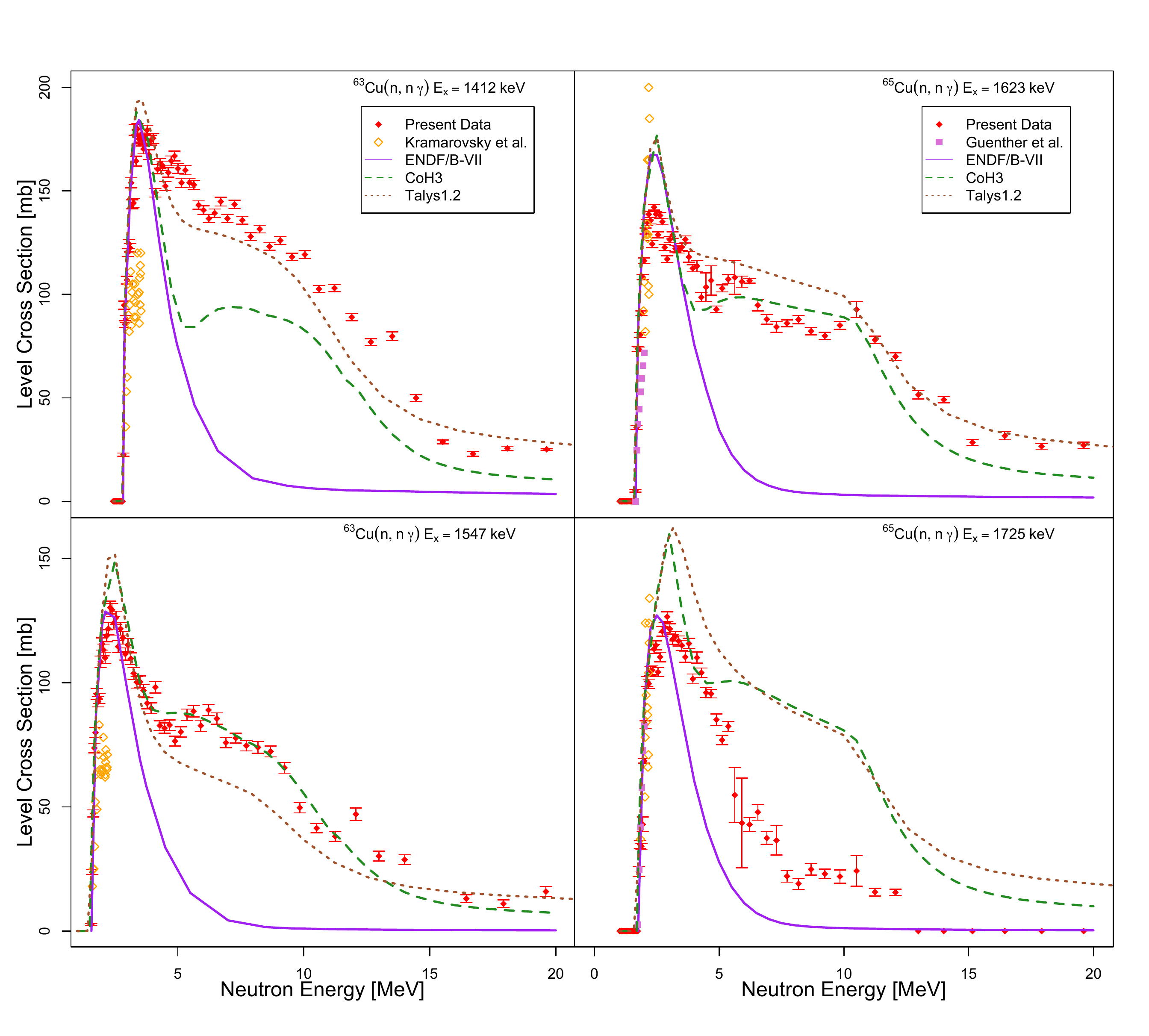}
\caption{The total $\gamma$-ray production cross section for the fourth (E$_x$=1412 \& 1623 keV) and fifth (E$_x$=1547 \& 1725 keV) energy levels in \tcu (left column) and \fcu (right column) from this work as well as the inelastic neutron scattering cross sections from  \cite{kram89,guenther}, and the ENDF/B-VII evaluation for these levels \cite{endf}.  The level de-excitation cross sections from the TALYS1.2 and CoH3 calculations are presented as well (dashed lines) for these level. For a description of the types of level de-excitation cross sections presented here please see Figure \ref{fig:ana:63cu669}.}
\label{fig:ana:cues}
\end{center}
\end{figure*}

Fig. \ref{fig:ana:cues} shows our experimental data for the fourth and fifth energy levels in both nuclei, together with previous measurements, as well as the ENDF/B-VII evaluation for the four levels shown.  The cross sections for the energy levels are determined by summing the $\gamma$ production cross sections for each of the decay branches.  In the case of the 1326-keV  level of \tcu, the level de-excitation cross section is determined by summing  the $\gamma$ production cross sections for the 1412, 742, and 442-keV \gray transitions.  The level de-excitation cross section for the 1623-keV level in \fcu was obtained by summing the $\gamma$ production cross sections of the 1623, 852.7, and 507.9 keV transitions.  The 1547-keV level de-excitation cross section was obtained by summing  the $\gamma$ production cross sections for the 1547, 877, and 584-keV \gray transitions.  Finally the 1725-keV level de-excitation cross section in \fcu was determined by summing  the $\gamma$ production cross sections for  1725.92, 954.5, and 609.5-keV \gray transitions.  The branching ratios for each of these energy levels was calculated from the ratio of the $\gamma$ production cross section for each transition to the total level de-excitation cross section.  In all cases the branching ratios for the transitions were compared with those given in ENDSF and were generally found to be in agreement with the literature. Branching ratios were computed for all 81 neutron energy bins used in this analysis, and were constant across the entire neutron energy range.  In general, the measured level de-excitation cross sections were in good agreement with previous experimental data, and for a limited energy-range with the ENDF/B-VII evaluation.

	 Guenther {\em{et. al}} measured the $\gamma$ production cross sections at two angles (55$^0$ and 125$^0$) with incident neutrons up to 4.5 MeV, and \cite{guenther}.   We find that our measurements, which include a much larger solid angle,  are about 10\% higher, although both measurements have comparable shapes.   Similar measurements for the first excited level in \fcu do not show such a disagreement.  Our measurements are consistent with $\gamma$ production cross section measurements from   Nishimura {\em{et. al}}\cite{nishimura} at the JAERI 5.5 MV Van De Graaff, as well as the measurements of Tucker {\em{et. al}} \cite{tucker} for \tcu.

\begin{figure}[htbp]
\begin{center}
\includegraphics[width=\linewidth]{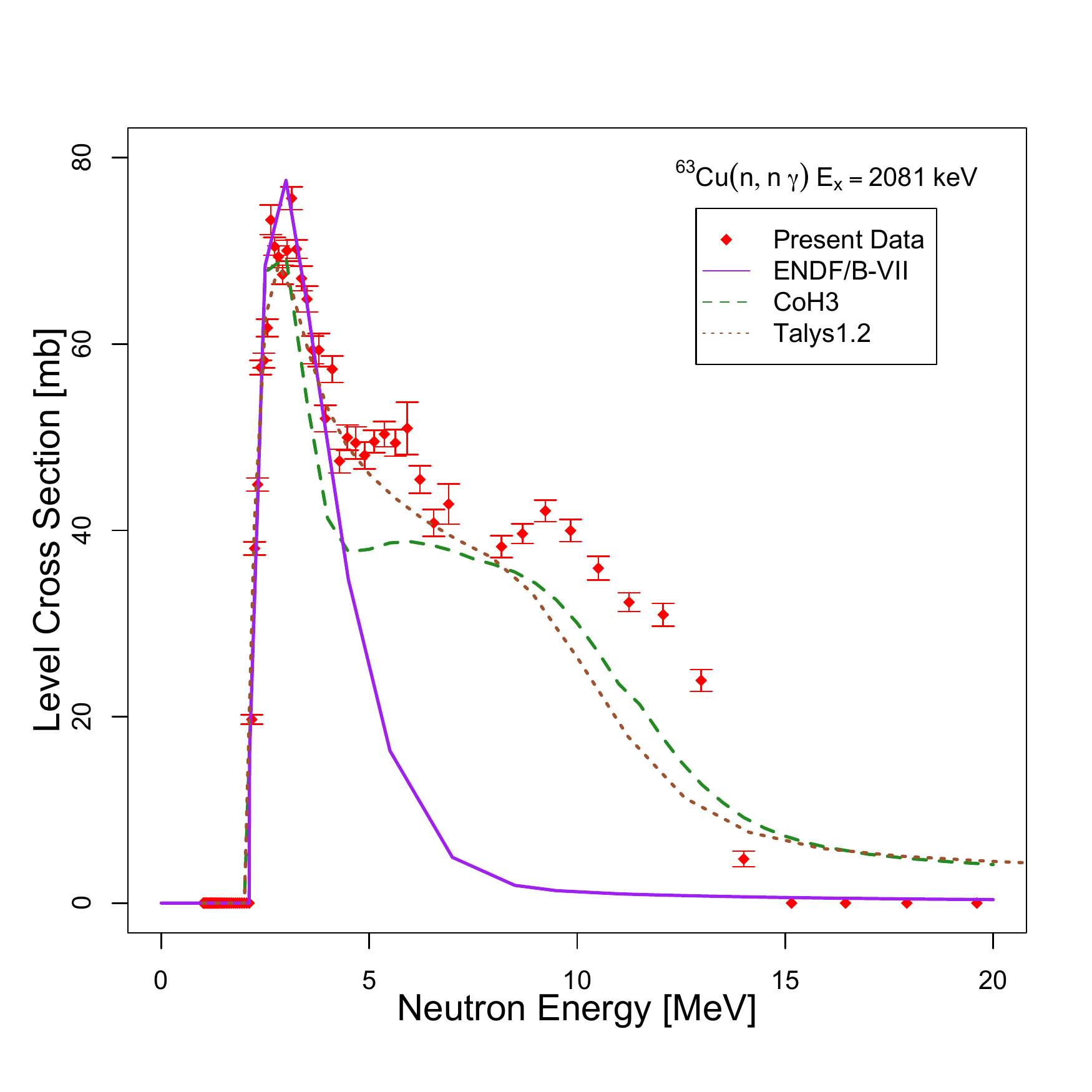}
\caption{The total $\gamma$-ray production cross section for the 2081-keV level in \tcu from this measurement, as well as the ENDF/B-VII evaluation for this excited level.  For a description of the types of level de-excitation cross sections presented here please see Figure \ref{fig:ana:63cu669}. The decay of this level produces a $\gamma$ ray in the vicinity of the $^{76}$Ge endpoint.}
\label{fig:ana:2081}
\end{center}
\end{figure}

Figure \ref{fig:ana:cues} shows the measured level de-excitation cross sections along with the available data, and the ENDF/B-VII evaluated level excitation cross sections for the fourth and fifth energy levels in both nuclei.  Our data agree with the limited data from measurements.  The lack of experiment data is important because these partial neutron cross sections typically provide the data for the ENDF evaluations that give the discrete level excitation cross sections.   Due to the difficulty of resolving closely spaced levels, neutron measurements, are usually available at only low energy levels and for a limited range of incident neutron energies.
Prior to our work, the ENDF/B-VII evaluation on Cu used very little experimental data above the fifth energy levels in either nuclei, and only included data for three incident neutron energies.  

\begin{figure}[htbp]
\begin{center}
\includegraphics[width=\linewidth]{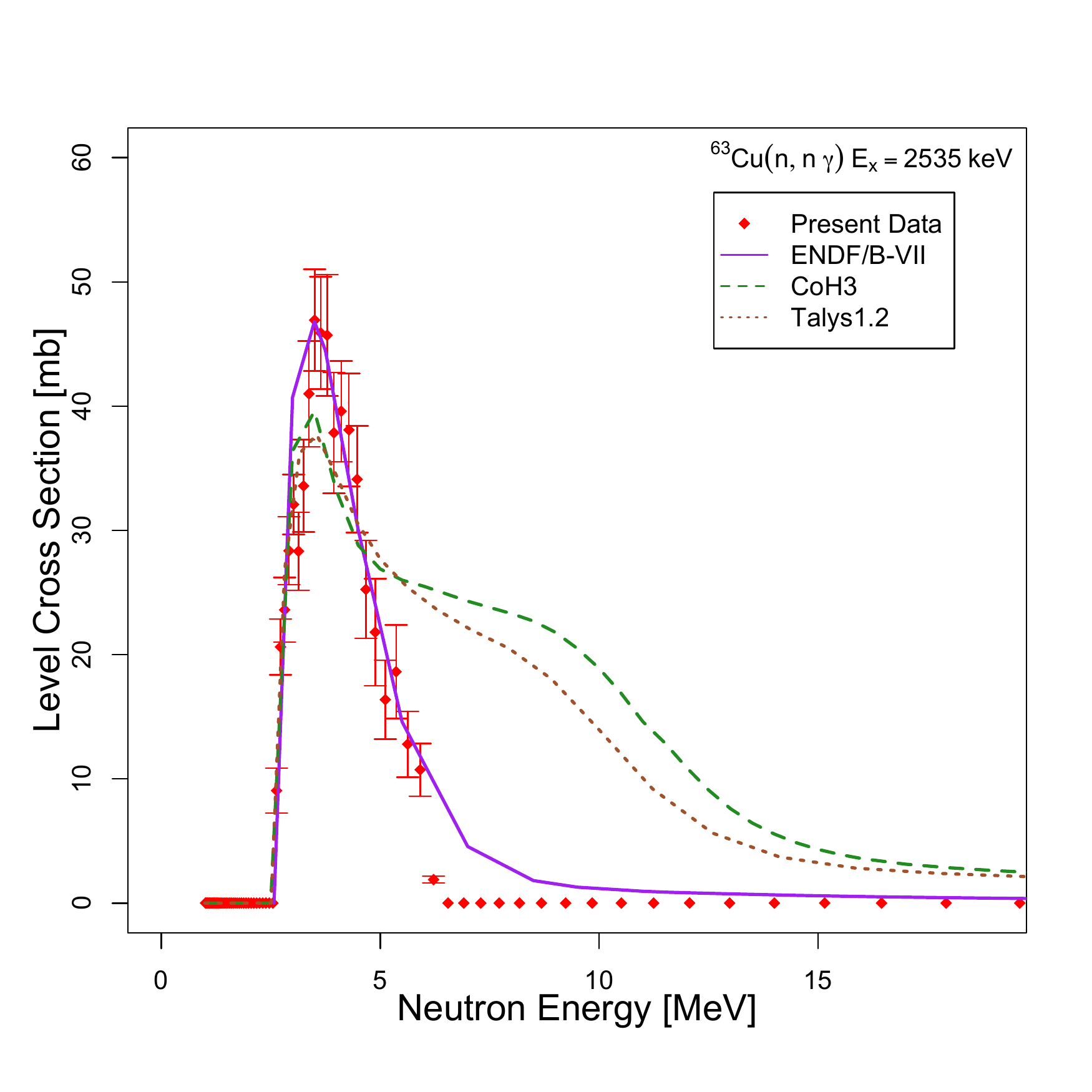}
\caption{The total $\gamma$-ray production cross section for the 2535-keV level in \tcu from this measurement, as well as the ENDF/B-VII evaluation for this level.  For a description of the types of level de-excitation cross sections presented here please see Figure \ref{fig:ana:63cu669}.  The 2535-keV level is in the vicinity of the $^{130}$Te endpoint energy. }
\label{fig:ana:tefep}
\end{center}
\end{figure}

\begin{figure}[htbp]
\begin{center}
\includegraphics[width=\linewidth]{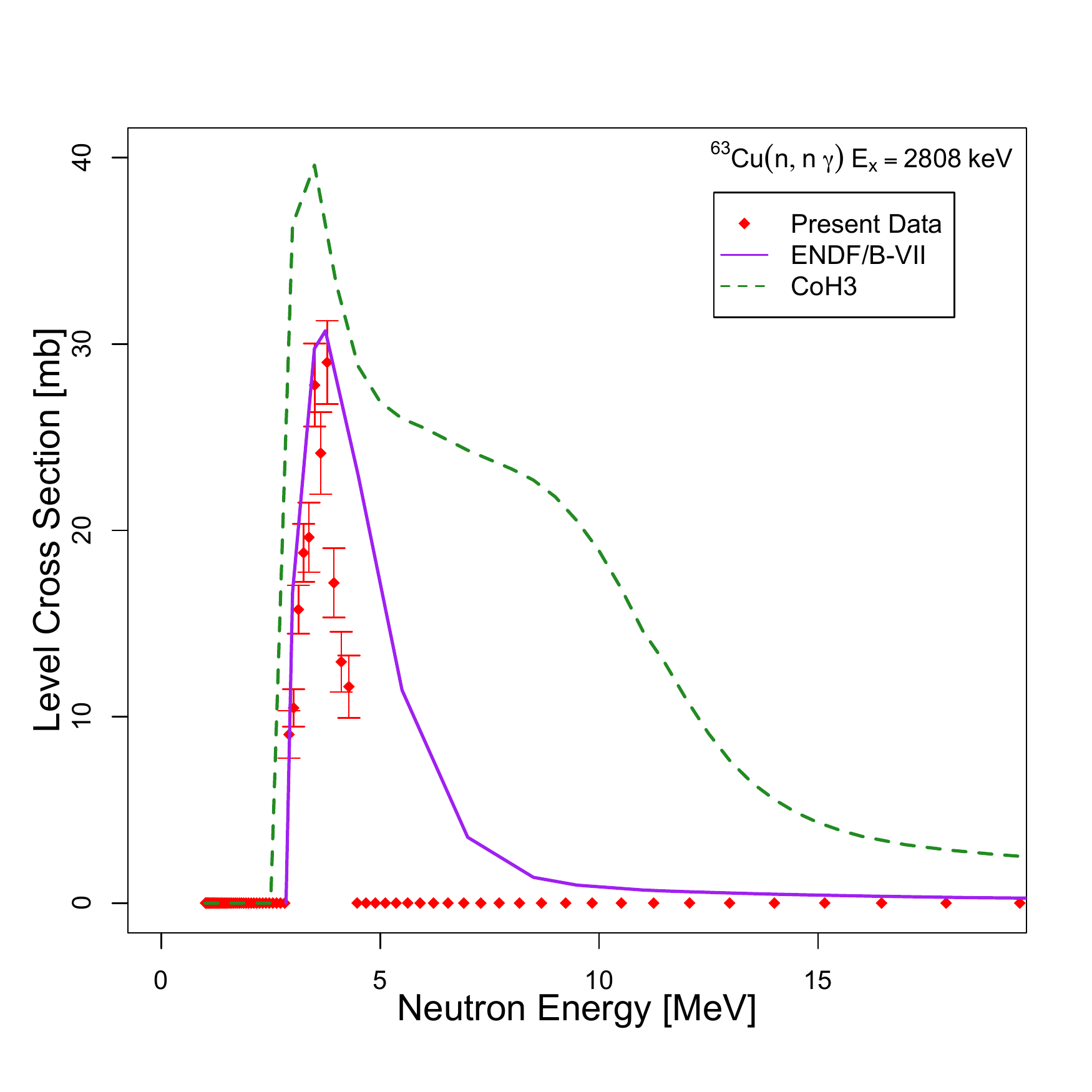}
\caption{The total $\gamma$-ray production cross section for the 2808-keV level in \tcu from this measurement, as well as the ENDF/B-VII evaluation for this level.  For a description of the types of level de-excitation cross sections presented here please see Figure \ref{fig:ana:63cu669}.  The 2808-keV level is in the vicinity of the $^{116}$Cd endpoint energy.  The range of discrete levels in the Talys1.2 simulation did not extend high enough in energy to provide a level de-excitation cross section for this state.}
\label{fig:ana:cdfep}
\end{center}
\end{figure}

	Figs \ref{fig:ana:2081}- \ref{fig:ana:cdfep} are a comparison of our measured level de-excitation cross sections in this experiment with the level excitation cross sections recommended by ENDF/B-VII where there was no experimental data available at the time of the evaluation.   Feeding from \grays at higher levels tends to have pronounced effects around 4-5 MeV, well above the peak cross section for the states discussed in this paper.  In the energy region at or below the maximum cross section, the population mechanism for the levels is assumed to be direct excitation and therefore is directly comparable with the ENDF/B-VII recommended level excitation cross sections.  Comparing the maximum cross sections of these levels with our experimental data reveals the issue with using the evaluation at higher energies;  the ENDF/B-VII evaluated discrete level cross sections at higher energy are typically determined by statistical models, and at least in this evaluation many of the evaluated discrete level cross sections disagreed with the experimental cross sections by more than 10\%.  
	
	   Of  the three levels presented here only one of the levels was in agreement with the ENDF/B-VII evaluation.  The decay of these energy levels produces $\gamma$ rays that are in the vicinity of the expected neutrinoless double beta decay signal for several neutrinoless double beta decay isotopes.    In the case of 2081-keV level in \tcu, the ENDF/B-VII evaluation overestimates the maximum cross section by about 25\%.  This level shows significant strength at the higher energies from feeding from higher-lying levels.  The evaluation, however, does not directly include the feeding effects for this level because the evaluation does not include discrete higher lying levels that directly feed this level.  In most simulation packages invoking the ENDF/B-VII data set, the feeding of this level is handled predominantly by the statistical models.   These models typically don't cascade through the defined levels, but rather through a statistically determined level density.  The 2535-keV level shows much better agreement with the ENDF/B-VII evaluation, the maximum cross sections are consistent with the measured data, and  because there is not significant feeding from higher lying states the shape of the level cross sections are in agreement with each other.    The maximum cross section of the 2808-keV level in \tcu, however is about 50\% lower than the ENDF/B-VII evaluation.  Overall, our experimental data found disagreement in 13 of the 24 ENDF/B-VII evaluated levels above 2 MeV: eight had maximum cross sections lower than the ENDF/B-VII estimate, three had maximum cross sections values higher than the evaluation, and two of the levels were not seen.   In all cases the discrepancies were larger than 10\%, and in one particular case, the ENDF/B-VII the maximum cross section was overestimated by a factor of three.

\begin{figure}[htbp]
\begin{center}
\includegraphics[width=\linewidth]{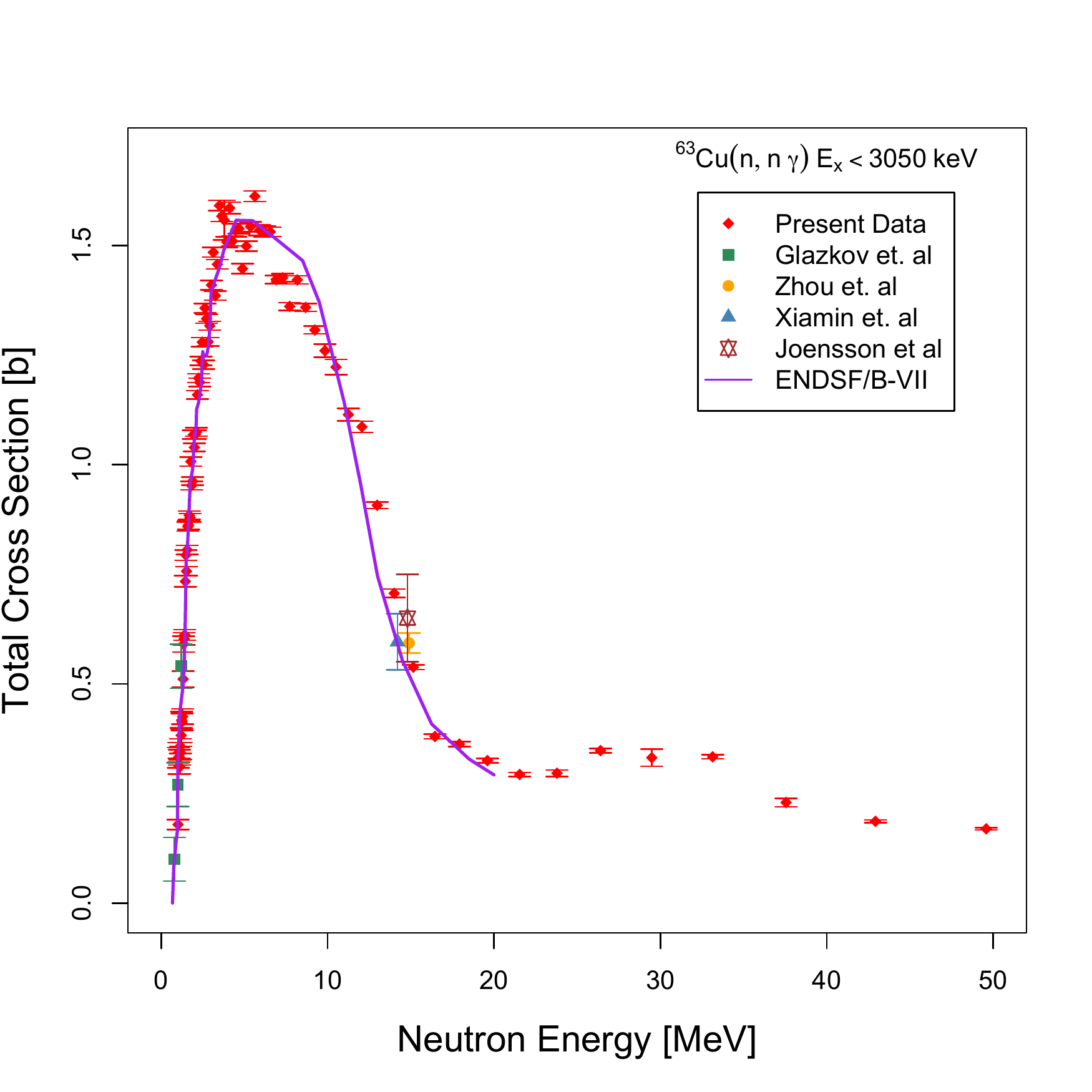}
\caption{The total neutron inelastic cross section for \tcu, as well as the results of (n,n') measurements\cite{glazkov} (squares), the \nnprime\ measurement \cite{zhou,xiamin} (circles), and the ENDF/B-VII evaluation for the total neutron inelastic cross section.  }
\label{fig:ana:63}
\end{center}
\end{figure}

\begin{figure}[tbp]
\begin{center}
\includegraphics[width=\linewidth]{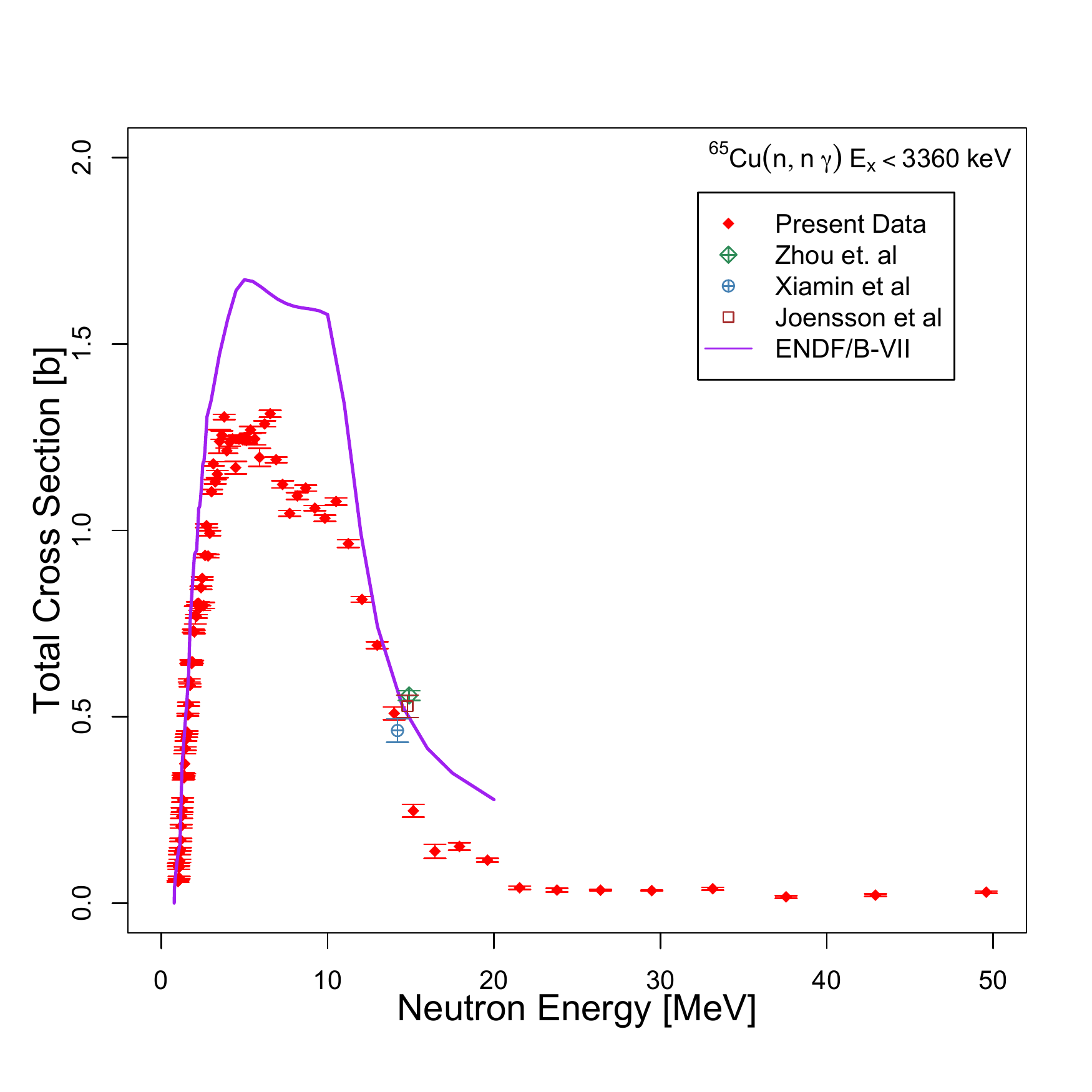}
\caption{The total neutron inelastic cross section for \fcu, as well as the results of an (n,n') measurements\cite{joensson} (squares), the \nnprime\ measurements \cite{zhou,xiamin} (circles), and the ENDF/B-VII evaluation for the total neutron inelastic cross section.  }
\label{fig:ana:65}
\end{center}
\end{figure}

	 In addition to comparing our measured level de-excitation cross sections with the ENDF/B-VII evaluated level excitation cross sections, we also compared our data with the ENDF/B-VII evaluated total neutron inelastic cross sections for \tcu and \fcu.   The total neutron inelastic cross sections were obtained by summing the $\gamma$ production cross sections for the  ground state transitions in each isotope, and are shown in Figs. \ref{fig:ana:63} and \ref{fig:ana:65}. The total inelastic cross sections deduced from $\gamma$ production cross sections are comparable to the ENDF/B-VII evaluation provided there isn't significant strength associated with the ground state decays of energy levels greater than 4 MeV.  In the case of \tcu, the total neutron inelastic cross section is in excellent agreement with both experimental data \cite{glazkov,zhou, xiamin,joensson}, and the ENDF/B-VII evaluation.  Around 20 MeV there is a slight increase in total neutron inelastic cross section from this measurement which is due to the \fcu(n,3n$\gamma$) reaction.  According to the description of the ENDF/B-VII evaluation, 26 levels below 3.7 MeV were used to build the level information for $^{63}$Cu \cite{cuendf}.  The levels used in the total neutron inelastic cross sections are the same below 2.889 MeV.  The ENDF/B-VII evaluation, however includes three levels at 3.3, 3.48, and 3.7 which are assumed to be collective excitations and provide significant contributions to the inelastic-scattering and \gray production cross sections \cite{cuendf}.   There was no evidence for any ground-state transitions from these levels, nor transitions to any of the levels assumed in the decay paths provided in the ENDF/B-VII evaluation.  It is difficult to evaluate the existence of these levels, since the exact excitation energies, spins, parities and decay paths are unknown.    There have been several inelastic proton experiments that have seen  levels at these energies with quite considerable strength \cite{iwa79,mcc66}, however there are no \gray  transitions in the current experimental data with sufficient strength to indicate the presence of these levels.  

	The total neutron inelastic cross section for \fcu (see  Fig. \ref{fig:ana:65}) is considerably lower than the ENDF/B-VII evaluation.  For this evaluation, the $\gamma$ production cross section for 19 levels below 3.360 MeV were included in the summation.  As with the \tcu evaluation, four collective levels were included in the ENDF/B-VII evaluation at 2533, 3080, 3350, and 3500 keV.  In addition to these three collective levels, a forth presumed collective level at 2533 keV was also not observed in our work.   The ground-state transitions were also not observed for these levels, nor was there any enhanced \gray transition that would indicate the existence of these strong collective excitations.  The total neutron inelastic cross section from this experiment is in agreement with the measurements of \cite{zhou}, but are somewhat lower than the results of \cite{xiamin}, and \cite{joensson}.  In general, the total neutron inelastic cross section for \fcu is about 30\% lower than the ENDF/B-VII evaluation, and the \cite{xiamin,joensson} measurements.  

	From the current measurement, there appears to be a significant discrepancy between the high-energy states in these two nuclei.   There are several potential explanations for the deficit in the continuum strength of \fcu  It is quite likely that the experimental value is low because of unobserved gamma transitions. Either the gammas are at higher energies, above the 4 MeV cutoff of the GEANIE spectra, or the decay is fragmented into many weaker gammas below the detection limits, or both. A long-lived isomer in \fcu could also remove some of the cross section, because GEANIE measures mostly prompt gammas, but no such isomers are known in \fcu.  The most likely explanation for the decrease in the continuum strength is unobserved $\gamma$ transitions.  Such unobserved strength is common in odd-odd nuclei close to the mid-shell; nuclei near the mid-shell region typically experience maximal deformation \cite{cast90}.

 \subsection{Comparison with Statistical Models}

 	Two different codes were used for the statistical model calculations in this analysis, TALYS-1.2\cite{talys}, and COH3\cite{kawano}.   These codes are both based on the Hauser-Feshback formalism, and both rely on local optical model parameters for the calculations.    The statistical models were used to produce $\gamma$ production cross sections for several transitions in \tcu and \fcu.  These $\gamma$ production cross sections were summed to give level de-excitation cross sections similar to those presented throughout the text.  The results of these calculations were compared with the first, fourth and fifth energy levels in both \tcu and \fcu, as well as the 2081-keV level in \tcu.    
	
	Examining the agreement between the various models and the experimental data, it becomes very apparent that neither model is able to  predict the strength of individual levels to within 10\% of the experimental data.  For the first energy levels in both Cu nuclei, both models provide a reasonable approximation of the actual cross section up to the maximum cross section.  The CoH3 calculations overpredict the effects of feeding from higher lying energy levels to the first excited level.  In fact for \tcu the CoH3 overpredicts the cross section by about 20\% between 5 and 15 MeV, while the TALYS-1.2 model underpredicts the cross section by about 30\% in the same energy region.  While for \fcu, both models overpredict the cross section in this region by about 55\%.   Overall, both codes overpredict the summed strength in \fcu by about 50\%, while for \tcu the summed strength for the CoH3 model is about 11\% higher than our data, and the TALYS-1.2 calculation is about 25\% lower than our data.  
	
	At higher energy levels there are similar problems.  Fig. \ref{fig:ana:cues} shows the third and fourth excited levels for both Cu nuclei.   The maximum cross sections for the third and fourth excited levels are overpredicted by both calculations for both nuclei.  The fourth and fifth excited levels in \tcu are overpredicted by 6-8\%, and 14-17\%, respectively.  While the same levels in \fcu are overpredicted by 25\% in both calculations.  The CoH3 model achieves reasonable agreement with the fourth excited level in \tcu, but has 67\% less total strength than our experimental data for the third excited level.  The TALYS-1.2 calculation underpredicts the summed strength in both excited levels:  having roughly 90\% of the observed total strength in the third excited level, and 85\% of the observed total strength in the fourth excited level.  The third excited level in \fcu is in reasonable agreement with both models; the CoH3 calculation slightly underpredicts the summed strength by about 10\%, while the TALYS-1.2 calculation overpredicts the summed strength by 11\%.  Neither calculation is in agreement with the fourth excited level in \fcu; both overestimate the summed strength by 260-290\%.  Finally, we turn to the 2081-keV level in \tcu.  This level is important for experiments like \majorana because the primary transition emits a \gray in the vicinity of the $^{76}$Ge endpoint energy.  For this level the calculations overpredict the maximum cross section by 6-8\%.  The TALYS-1.2 calculation underpredicts the strength in the continuum region (taken to be between 5-15 MeV) by about 6\%, while the CoH3 calculation is roughly in agreement with the experimental data.

	 The statistical models typically in use for estimating $\gamma$ production cross sections for individual excited levels display inconsistencies  from nucleus to nucleus, and even among excited levels in the same nucleus.  Neither model seems to be able to accurately predict the strength in a particular level to the precision necessary for double-beta decay and other high-sensitivity, low-background experiments.   The models vary considerably in their modeling of a particular level; showing agreement to within 10\% in one level and then over predicting the summed strength in a neighboring level by a factor of 3.  In addition, both codes seem to overpredict the contributions from the continuum in \fcu.  Experiments designed to search for rare signals require a better understanding of their background contributions than can be afforded with the current statistical model calculations.

\subsection{Monte Carlo Simulations}
	For most $\beta\beta$ studies the successful observation of a $0\nu\beta\beta$ signal relies on the ability to achieve an extremely low-background experimental setup.  As this is the first generation of experiments to attempt such a low sensitivity, the estimates of the background levels are highly dependent on simulations.   To quantify the accuracy of these simulations, we compared our experimentally determined total inelastic cross sections with those extracted from simulations.  For our analysis we simulated ten-million neutrons scattering off a 139-g block of $^{nat}$Cu with the same dimensions as our actual target.  The neutrons had a flat energy distribution between 0 to 40 MeV.

\begin{figure}[tbp]
\begin{center}
\includegraphics[width=\linewidth]{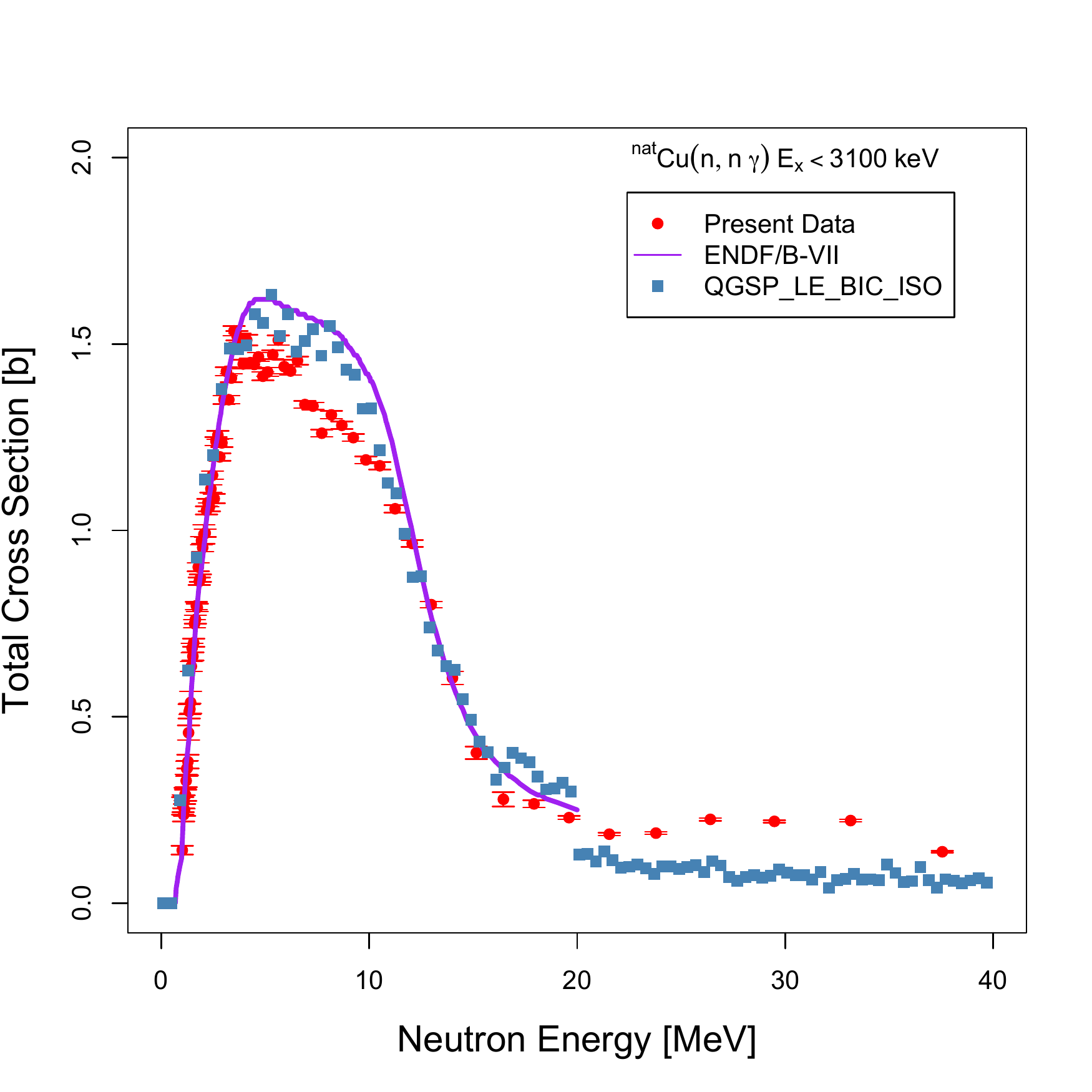}
\caption{The total inelastic cross section for $^{nat}$Cu (circles),  the ENDF/B-VII evaluation (line), and the results of the GEANT4 simulation (squares). }
\label{fig:ana:mc:nng}
\end{center}
\end{figure}

\begin{figure}[tbp]
\begin{center}
\includegraphics[width=\linewidth]{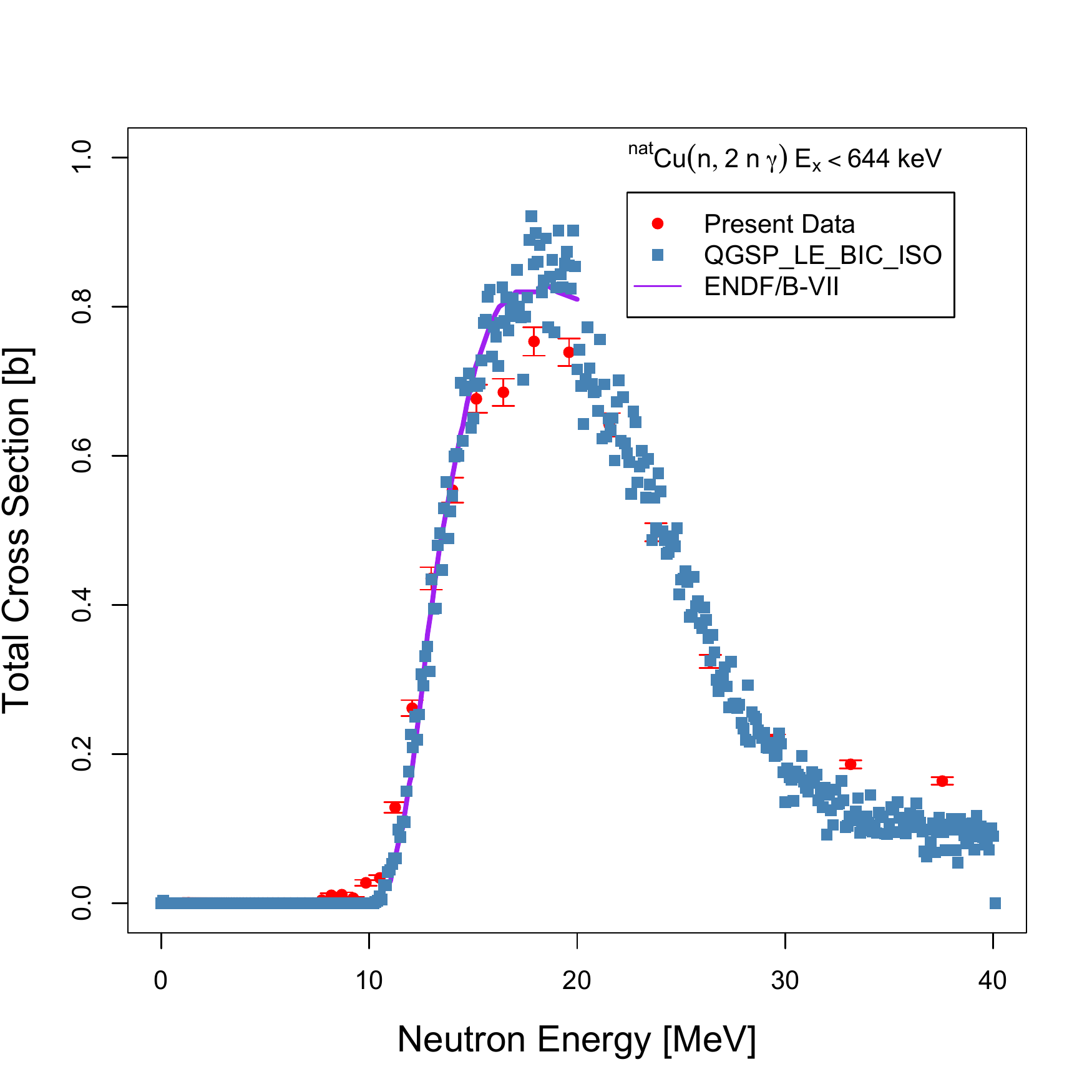}
\caption{The (n,2n) cross section for $^{nat}$Cu(circles),  the ENDF/B-VII evaluation (line), and the results of the GEANT4 simulation (squares). }
\label{fig:ana:mc:n2n}
\end{center}
\end{figure}

	The simulations were done in GEANT4-09-03 using the QGSP\_LE\_BIC\_ISO physics list for inelastic neutron scattering \cite{geant4,*geant42}.  The ENDF/B-VII libraries were called to account for both elastic and inelastic neutron scattering below 20 MeV.  Above 20 MeV, the elastic scattering process was modeled using the low-energy elastic model, while the inelastic process was modeled from 19.9 MeV to 9.9 GeV with the Binary Cascade model \cite{g4physics}.  The results of the simulation are shown in Figs. \ref{fig:ana:mc:nng} and \ref{fig:ana:mc:n2n}.    The curve marked as QGSP\_LE\_BIC\_ISO refers to the summation of all inelastic scattering processes ($n,n'\gamma$) that occur for a particular incident neutron, and normalized to the total number of simulated neutrons at that particular incident energy.   The red circles, denoted as the present data, are the total inelastic cross sections for $^{nat}$Cu as a function of incident neutron energy.  These data points represent the summed strength of 25 levels  and 14 levels  below 3.1 MeV in  \tcu and \fcu, respectively.   The ENDF/B-VII evaluation for \natcu is depicted by the solid purple line, and is the sum of the natural abundance corrected \tcu and \fcu total inelastic cross sections.   As can be seen from Figs. \ref{fig:ana:mc:nng} and \ref{fig:ana:mc:n2n}, GEANT4 simulations are consistent with the ENDF/B-VII evaluation.

\begin{figure}[tbp]
\begin{center}
\includegraphics[width=1.1\linewidth]{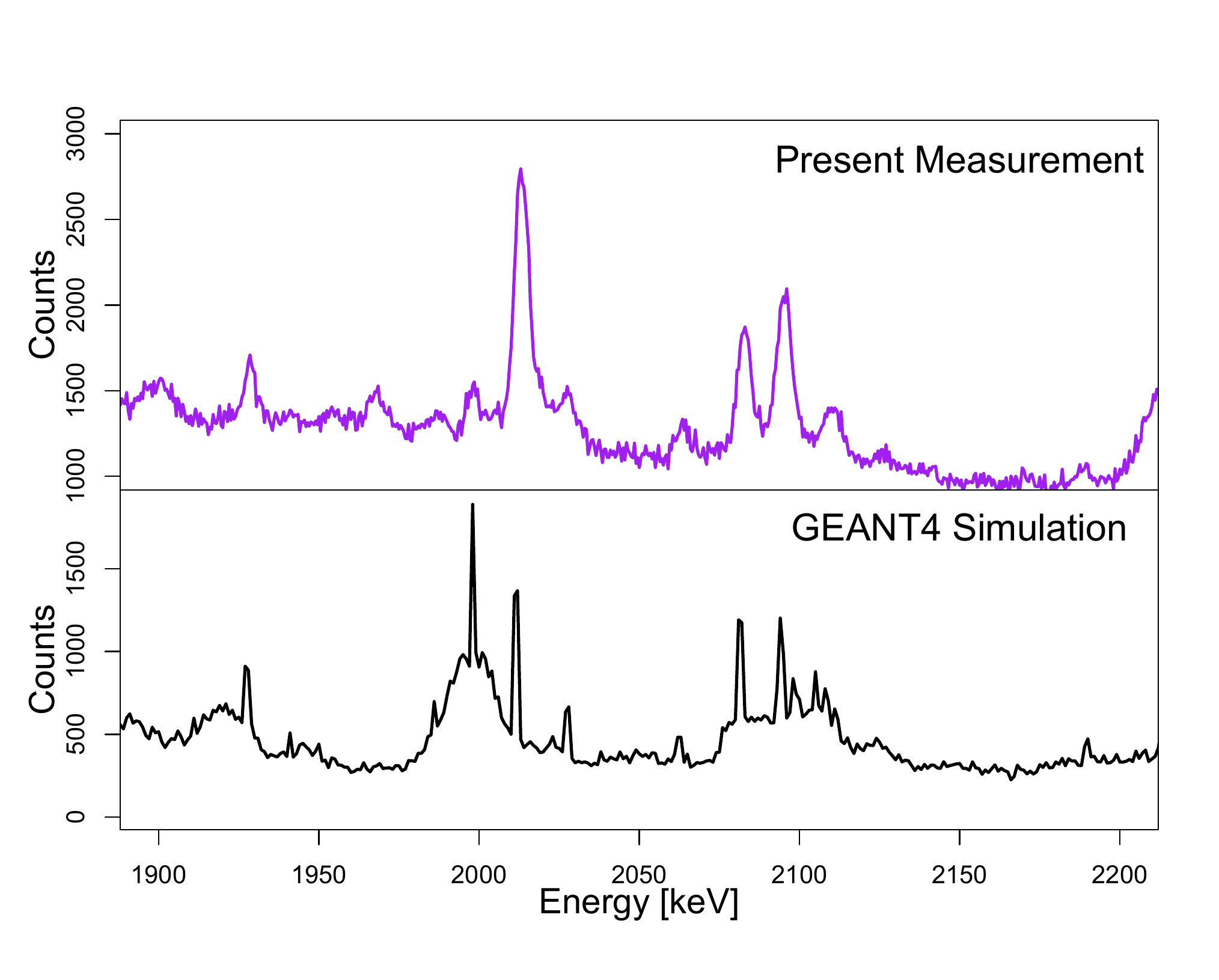}
\caption{Comparison of spectra produced from GEANT4 simulation with that obtained from the present experiment.  The top figures shows the experimental spectrum from this experiment, while the bottom spectrum shows the results of a GEANT4 simulation.  The specifics of the GEANT4 simulation are outlined in the texts.  }
\label{fig:ana:spect}
\end{center}
\end{figure}

	Geant4 implements the ENDF/B-VII evaluations in much the same way as other simulation programs.  Once the inelastic process is activated a random number generator is used to decide among the various available energy levels, which are weighted according to their contribution to the total cross section.	In the region where the ENDF/B-VII provides level excitation cross sections, and branching ratios,  the simulations produce \gam\ rays that correspond with known levels and branching ratios in the particular nucleus.  The issue comes when the excitation energy exceeds the ENDF/B-VII cutoff energy for discrete level, and the ENDF/B-VII continuum file is employed.  The summed strength in this file usually accounts for upwards of 70\% of the total strength in the ENDF/B-VII evaluation, and unlike the evaluated level data, doesn't have a decay path, rather it relies on model-generated $\gamma$ production probabilities, gamma-ray emission spectra, and neutron emission-energy spectra.  At this point the decay no longer proceeds through defined levels in the nucleus, and the emitted \gam\ rays no longer correspond to transitions between nuclear levels.   In addition given the significant strength associated with this file, this region becomes significant at very low incident neutron energies; neutron above 3 MeV interacting in Cu are predominately described by the ENDF/B-VII continuum file.  Figure \ref{fig:ana:spect} shows a comparison of the GEANT4 simulated spectrum compared with the experimental spectrum observed in the present experiment.  The simulation shows significant strength in the 2 MeV region, which is not observed in the actual experimental spectrum.  The ENDF/B-VII evaluation appears to reproduce some \gray transitions very nicely.  Although there appears to be some strange structure that appears below many peaks that is not seen in the actual experiment data (see for example the 2 and 2.1 MeV regions in Fig. \ref{fig:ana:spect}).  Furthermore, the simulation predicts strength from more states than are actually seen in the data, while also the relative strengths of several transitions do not appear to agree with the actual data.    

\subsection{Cross Sections relevant to $0\nu\beta\beta$ decay searches }
	Cu is a very popular material in $0\nu\beta\beta$ decay searches.  It has good thermal properties which allows its use in cooling systems.  Cu can also be manufactured extremely cleanly, and thus is generally used in large quantities around these low-background setups.  One issue that has not been examined is the probability that neutrons interacting in Cu will produce $\gamma$ rays that might interfere with, or even replicate a $0\nu\beta\beta$ signal.  Only four of the nine potential $0\nu\beta\beta$ isotopes listed in Table \ref{ana:tab:fep} had Q-values below the cutoff for ENDF/B-VII evaluated excited levels in either Cu nucleus.  The strength associated with excited levels above the ENDF/B-VII cutoff energy have been folded into the continuum file, and thus a proper parameterization of the population, and decay of these excited levels will not be explicitly provided to simulations employing ENDF/B-VII.

	A quick literature search revealed several $\gamma$ ray lines that could potentially interfere with several isotopes frequently used in $\beta\beta$ studies (see Table \ref{ana:tab:fep}).  In all, 12 $\gamma$-ray lines were identified as potential background lines for the $0\nu\beta\beta$ experiments.  Of these 12 $\gamma$-ray lines, seven were outside of the chosen energy range of the GEANIE detectors for this experiment.  Three of these high-energy $\gamma$-ray lines had detectable lower-energy $\gamma$-ray transitions which allowed for a cross section estimate.   The energy regions surrounding the Q-value, Q-value + 511 keV (SEP, single-escape peak), and the Q-value + 1022 keV (DEP, double-escape peak) for the isotopes listed in Tab. \ref{ana:tab:fep} were examined to rule out any additional unknown $\gamma$-ray lines that might not be currently noted in the literature.  Obviously, the $\gamma$ ray lines with sufficient energy to produce DEP's near the Q$_{\beta\beta}$  of $^{82}$Se, $^{100}$Mo, and $^{150}$Nd were too high in energy to be examined in this study.

\begin{table}[htdp]
\caption{ A list of frequently studied  $\beta\beta$ isotopes and their Q-values.   $\gamma$-production cross sections or upper limits for important transitions in  $^{nat}$Cu are given for neutrons between  2.897 and 4.196 MeV.   Where the cross section is listed as NA, the experimenters were unable to place a limit due the \gray being outside the range of the detection system.\label{ana:tab:fep} } {
	\begin{tabular}{@{}cccc@{}}
	\hline
	$\beta\beta$ isotopes & $\gamma$ ray & SEP & DEP\\
	& [mb] & [mb] & [mb] \\
	\colrule
	$^{48}$Ca & \fcu  NA & \fcu  NA & \fcu  NA \\
	$^{76}$Ge & \fcu $<$0.388(3)   NA &   &  \\
	$^{82}$Se &  \tcu 9.42(32) & $<$0.324(3) & \tcu  NA\\
	$^{96}$Zr &  \fcu 1.12(2) & $<$0.241(3) & \fcu NA\\
	$^{100}$Mo &\tcu 9.42(32) & \tcu 0.59(22) & \tcu  NA\\
	$^{116}$Cd & \tcu 4.41(23) &  & \\
	$^{130}$Te & \tcu 9.42(32) & \tcu 9.42(32) & \tcu $<$0.316(3) \\
	$^{136}$Xe & \tcu 0.62(10) & $<$ 0.392(3) & \tcu 1.03(10)  \\
	$^{150}$Nd & $<$ 0.319(3)  & $<$ 0.265(3) & \fcu  NA\\
	\hline
	\end{tabular}
}
\end{table}%

\paragraph{$^{48}$Ca Endpoint Energy}
	The CANDLES experiment is an experiments  looking to study $0\nu\beta\beta$ in $^{48}$Ca \cite{candles}.  The endpoint energy for this isotope is above the energy range of the GEANIE spectrometer in this experiment, and thus cross section information is not available for this isotope.  Recent measurements have shown that CaMoO$_4$ scintillators might be a viable option for studying $^{48}$Ca; these types of detectors would potentially have a 4\% energy resolution at 3-MeV when operated at  244 $^o$K \cite{2005Bel}.  With this energy resolution,  there are 35 \gray transitions in Cu in a 340-keV energy region around the $^{48}$Ca endpoint energy.  All of these \gray transitions derive from excited levels that are  above the cutoff energy for the ENDF/B-VII evaluated discrete levels in either Cu nucleus.

\paragraph{$^{76}$Ge Endpoint Energy}
	Experiments designed to study neutrinoless double beta decay in $^{76}$Ge typically employ enriched Germanium diodes as an active detection system.  These crystals have excellent energy resolution in the endpoint region -- typically corresponding to a resolution of $\approx$ 0.2\% or about 4 keV \cite{2004Aal}.  The literature states that $^{65}$Cu has a high energy level ($E_x$ = 3157(3) keV) that emits a 2041(3)-keV $\gamma$-ray line during the decay.    The 2041-keV $\gamma$ ray line associated with the decay of this excited level  was not observed in the GEANIE data, and consequently only a limit can be placed on the $\gamma$ production cross section.   A review of the literature revealed that this is a weakly excited level, currently only detected in a limited number of experiments \cite{1987Do,2000KO}.   Fortunately this decay is not to the ground state, and should be accompanied by a 1115-keV $\gamma$-ray line.  An examination of the $\gamma$ rays emitted in coincidence with the 1115-keV $\gamma$-ray line did not yield any strength in the 2041-keV region.  Table \ref{ana:tab:fep} gives the sensitivity of our measurements to a $\gamma$ ray produced in this energy region.   If this excited level exists and it decays via a 2041-keV $\gamma$ ray, it would likely have a maximum $\gamma$ production cross section below 0.35 mb.   There are no additional \gray transitions in the SEP \& DEP energy region.

\begin{figure}[tbp]
\begin{center}
\includegraphics[width=\linewidth]{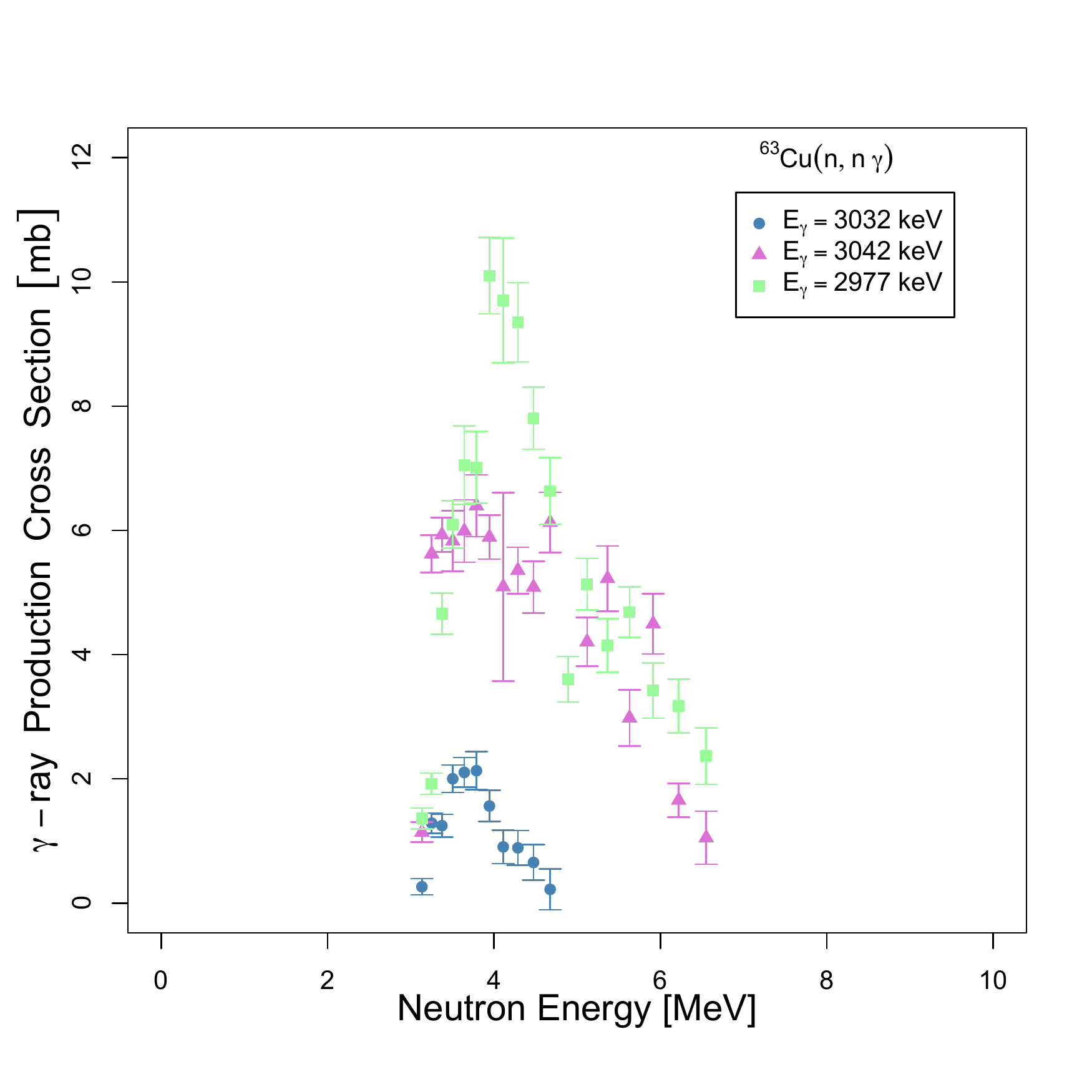}
\caption{The $\gamma$ production cross sections for 2977, 3032, and 3042-keV \grays in \tcu.   These \gray are all ground state transitions, and therefore the energy of the \gray corresponds to the energy level in \tcu.   These \gray transitions are in the vicinity of the $^{82}$Se and $^{100}$Mo endpoint energies.  The corresponding levels for these \grays are not included in the ENDF/B-VII evaluation for \tcu.}
\label{fig:ana:mofep}
\end{center}
\end{figure}

\begin{figure}[tbp]
\begin{center}
\includegraphics[width=\linewidth]{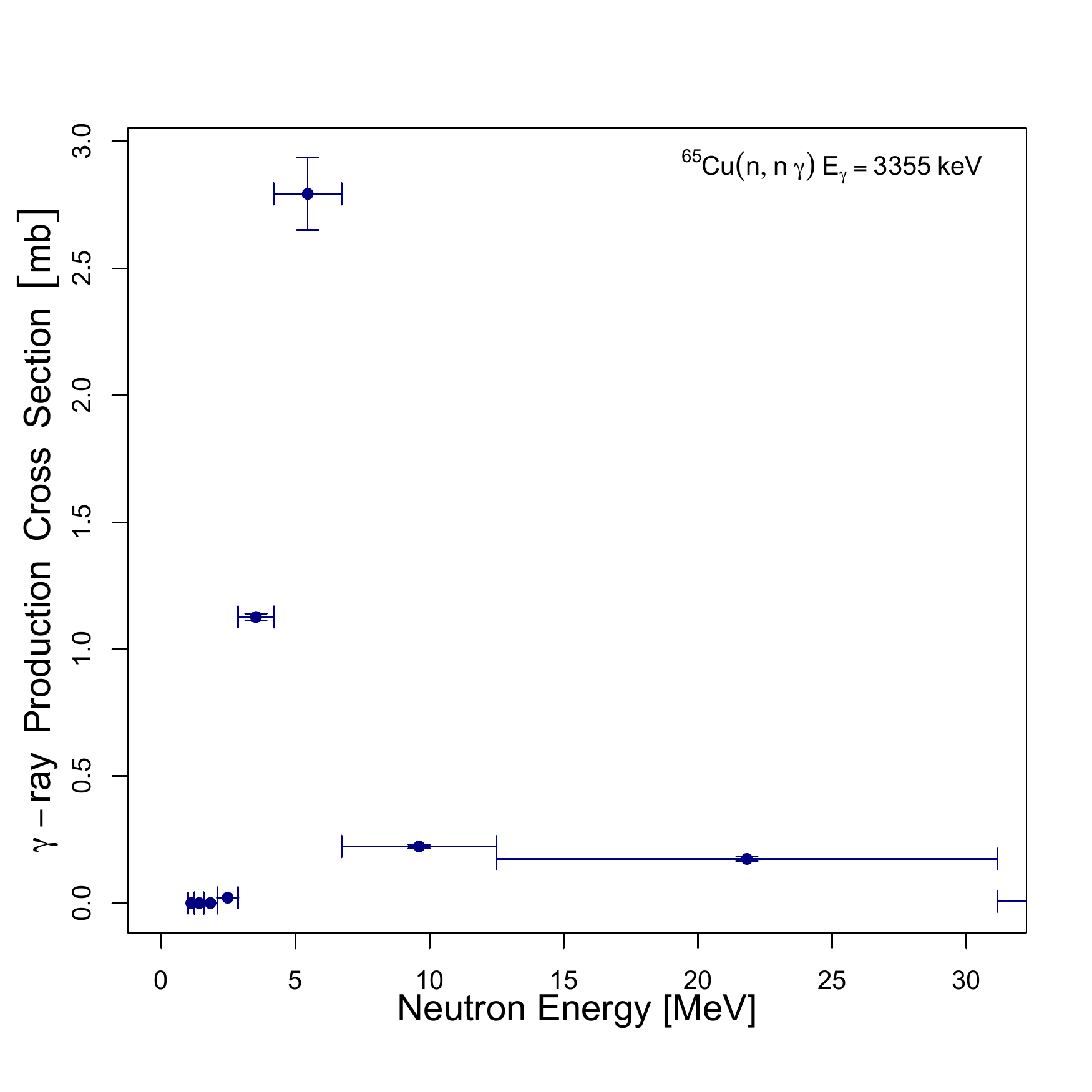}
\caption{The $\gamma$ production cross section for 3355-keV \gray in \fcu.  This $\gamma$ ray transition is in the vicinity of the $^{96}$Zr endpoint energy, and is a ground state transistion from the 3355 energy level in \fcu. The corresponding level for this \gray is not included in the ENDF/B-VII evaluation for \fcu.}
\label{fig:ana:zr}
\end{center}
\end{figure}

\paragraph{$^{82}$Se, $^{96}$Zr, and $^{100}$Mo Endpoint Energy}
	The Selenium, Zirconium and Molybdenum isotopes have all been grouped together because experiments such as MOON and SuperNEMO are designed to measure any of these isotopes, in addition to several other isotopes in their setups \cite{Moon,nemo}.  There doesn't appear to be many interfering $\gamma$ rays lines in the immediate vicinity of $^{82}$Se.   There are several excited levels that emit $\gamma$ rays on either side of the 2995-keV endpoint energy in $^{82}$Se, and the $\gamma$ production cross sections are shown in Fig. \ref{fig:ana:mofep}.  There is a single excited level that emits an interfering $\gamma$ ray within 5 keV of the $^{96}$Zr endpoint energy, although this decay path is not the dominant decay channel for this energy level.  The $\gamma$ production cross sections for this level, located at 3335 keV, is shown in Fig. \ref{fig:ana:zr}.  Both nuclei exhibit \gray lines that could double-escape into the respective regions of interest.   Again in the case of Zr, the decay channel that emits a 4371-keV $\gamma$ ray from the 7472.7-keV excited level is quite weak, occurring with a 12\% probability.   

		In the immediate vicinity of the $^{100}$Mo endpoint energy there is a ground-state transition from the  3032.70-keV energy level from $^{63}$Cu.  The $\gamma$ production cross section for this transition as a function of energy is shown in Fig. \ref{fig:ana:mofep}.  	 Expanding the search region to 3\% energy resolution reveals an additional 14 \gray transitions in $^{nat}$Cu  in the Mo region of interest.   A similar search in the SEP and DEP regions revealed an additional 34 potential background \gray lines.   Again, as in the case of Ca, the excited levels that produce these \gray lines are above the cutoff energy for the ENDF-BVII evaluated discrete levels in either Cu nucleus.

\begin{figure}[htbp]
\begin{center}
\includegraphics[width=\linewidth]{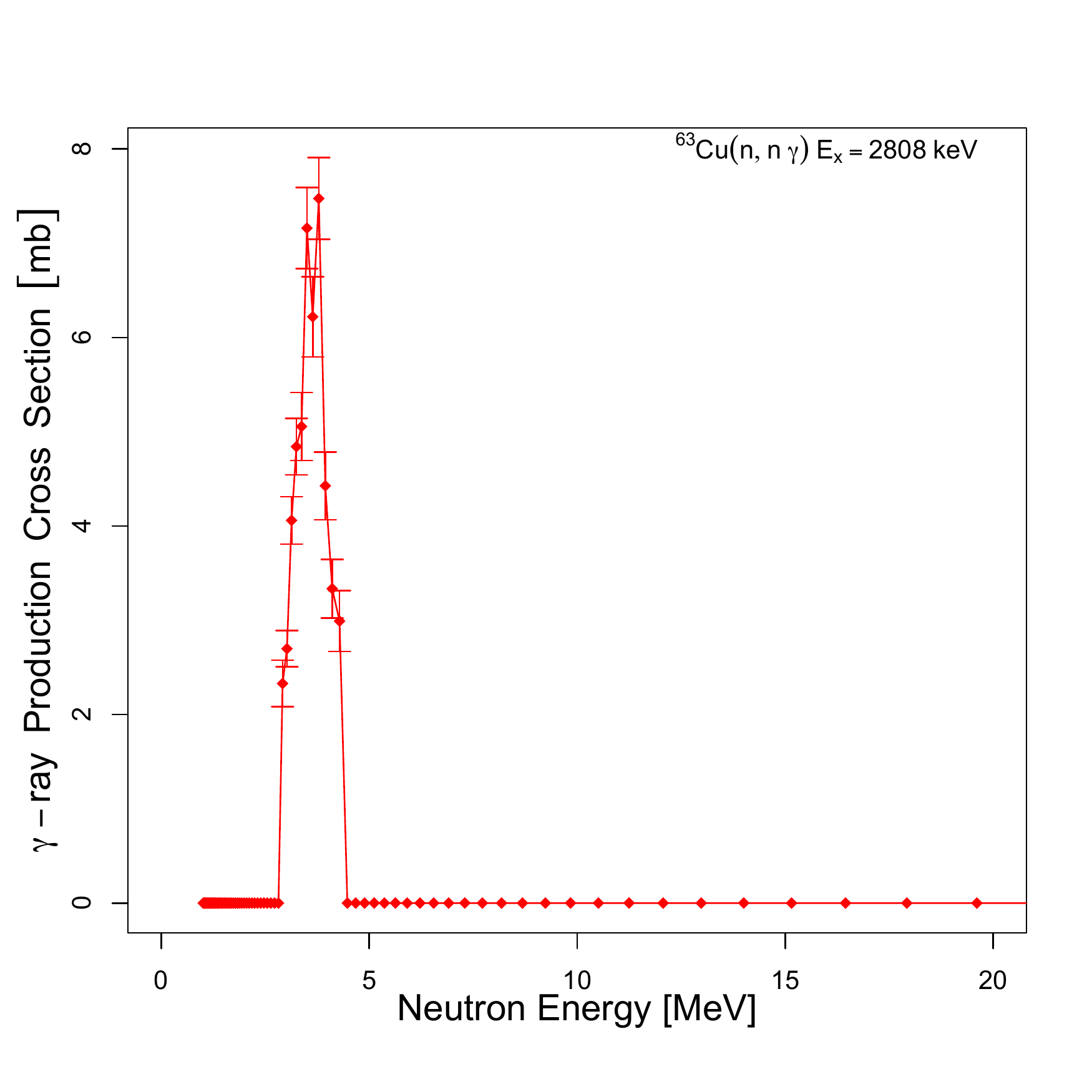}
\caption{The $\gamma$ production cross section for 2808-keV \gray in \tcu.  This $\gamma$ ray transition is in the vicinity of the $^{116}$Cd endpoint energy.  The corresponding level excitation cross section for this transition is shown in Fig. \ref{fig:ana:cdfep} and is included in the ENDF/B-VII evaluation for \tcu.}
\label{fig:ana:gCd}
\end{center}
\end{figure}

\paragraph{$^{116}$Cd Endpoint Energy}
	In the endpoint energy region of $^{116}$Cd there is a ground state transition from an excited energy level in $^{63}$Cu that is in the vicinity of the $^{116}$Cd endpoint energy.  While the dominate decay path for this level is to the ground state, there are additional decay branches that might facilitate the identification of this as a potential background line.    Fortunately, this level is included in the ENDF/B-VII evaluated discrete levels, and is shown in Fig. \ref{fig:ana:cdfep} together with the data from the present measurement.  The maximum cross section for this level is roughly 60\% higher in the ENDF/B-VII evaluation than in the present experimental data.   The $\gamma$ production cross section for this transition is shown in Fig. \ref{fig:ana:gCd}.  The typical energy resolution for CdZnTe is about 2\% at the Cd endpoint energy.  In such a large energy range, we can expect to see an additional 11, 12, and 5 levels producing potential background $\gamma$ rays in the full energy peak (FEP), SEP, and DEP regions, respectively.

\begin{figure}[htbp]
\begin{center}
\includegraphics[width=\linewidth]{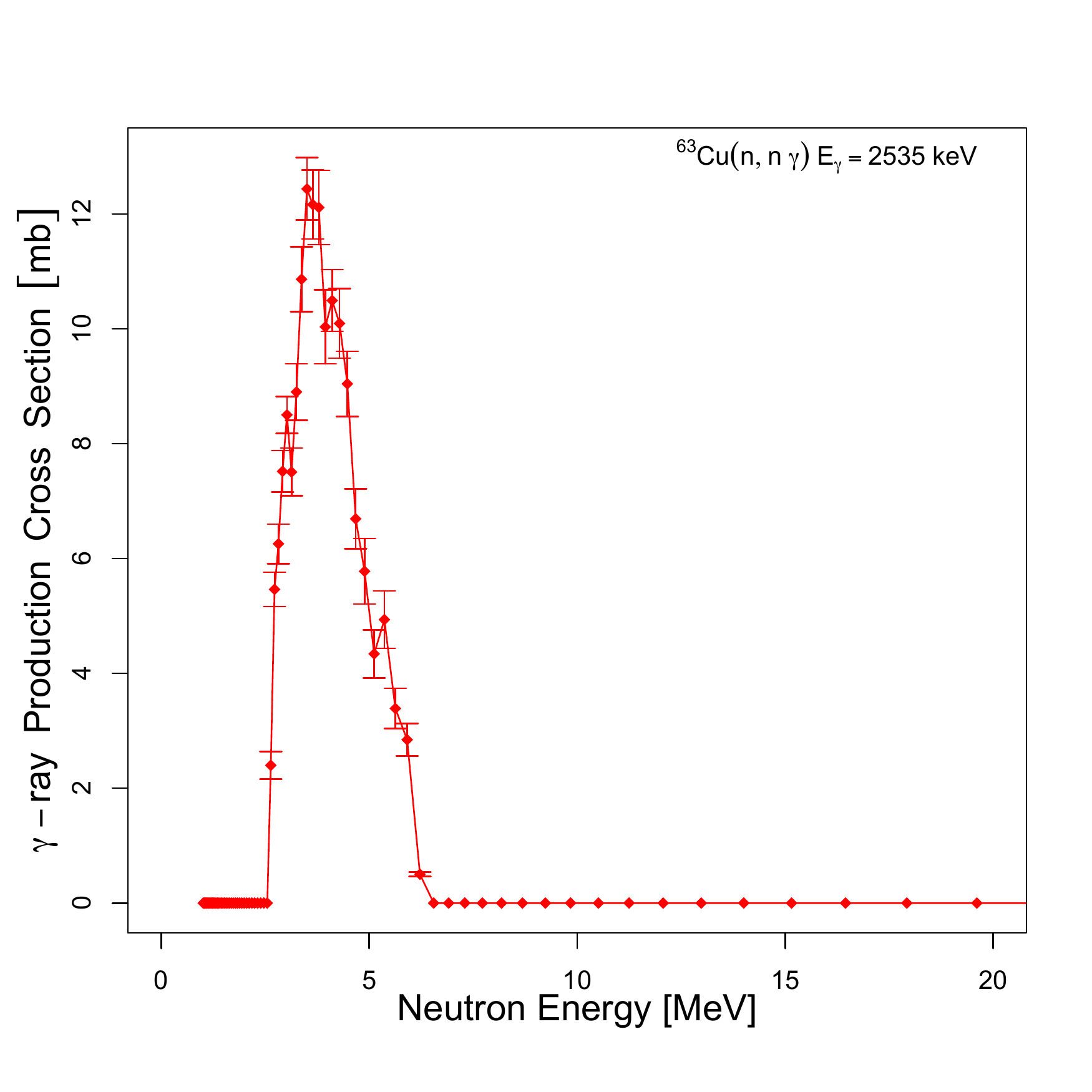}
\caption{The $\gamma$ production cross section for 2535-keV \gray in \tcu.  This $\gamma$ ray transition is in the vicinity of the $^{130}$Te endpoint energy.  The corresponding level excitation cross section for this transition is shown in Fig. \ref{fig:ana:tefep} and is included in the ENDF/B-VII evaluation for \tcu.}
\label{fig:ana:gte}
\end{center}
\end{figure}

\paragraph{$^{130}$Te Endpoint Energy}
 		The endpoint of the $^{130}$Te decay  coincides with a ground-state transition from the  2533-keV level in $^{63}$Cu; a level with significant strength (see Fig. \ref{fig:ana:tefep}).  The $\gamma$ production cross section for this transitions is shown in Fig. \ref{fig:ana:gte}.   This transition had the strongest $\gamma$ production cross section of any other transition in this energy region.  It is the strongest observed transitions above 1 MeV.  Current double beta decay measurements involving Te typically employ TeO$_2$ crystals used as bolometers.  There don't appear to be any additional troublesome \gray transitions from excited levels for this isotope, even when the energy window is expanded out to the anticipated 5-keV energy window.  The 2533-keV level has eight decay paths, with the dominate path being to the 1412-keV level in $^{63}$Cu.  For a discussion of potential backgrounds from the $^{130}$Te crystal please see Ref. \cite{dol11}.

\begin{figure}[tbp]
\begin{center}
\includegraphics[width=\linewidth]{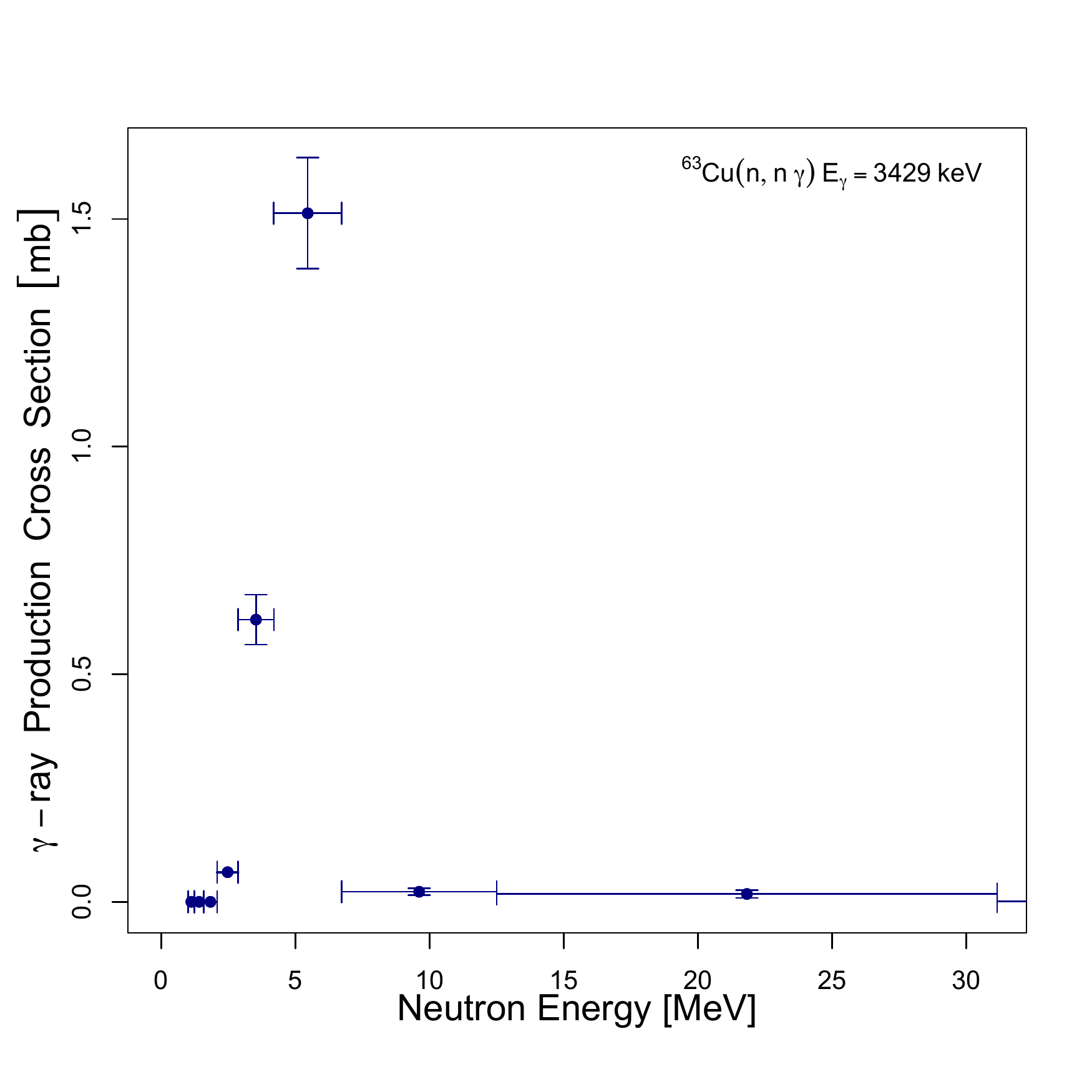}
\caption{The $\gamma$ production cross section for 3429-keV \gray in \tcu.  This $\gamma$ ray transition is in the vicinity of the $^{136}$Xe endpoint energy, and is a ground state transition from the 3429-keV energy level in \tcu. The corresponding level for this \gray is not included in the ENDF/B-VII evaluation for \tcu.}
\label{fig:ana:xe}
\end{center}
\end{figure}

\begin{figure}[tbp]
\begin{center}
\includegraphics[width=\linewidth]{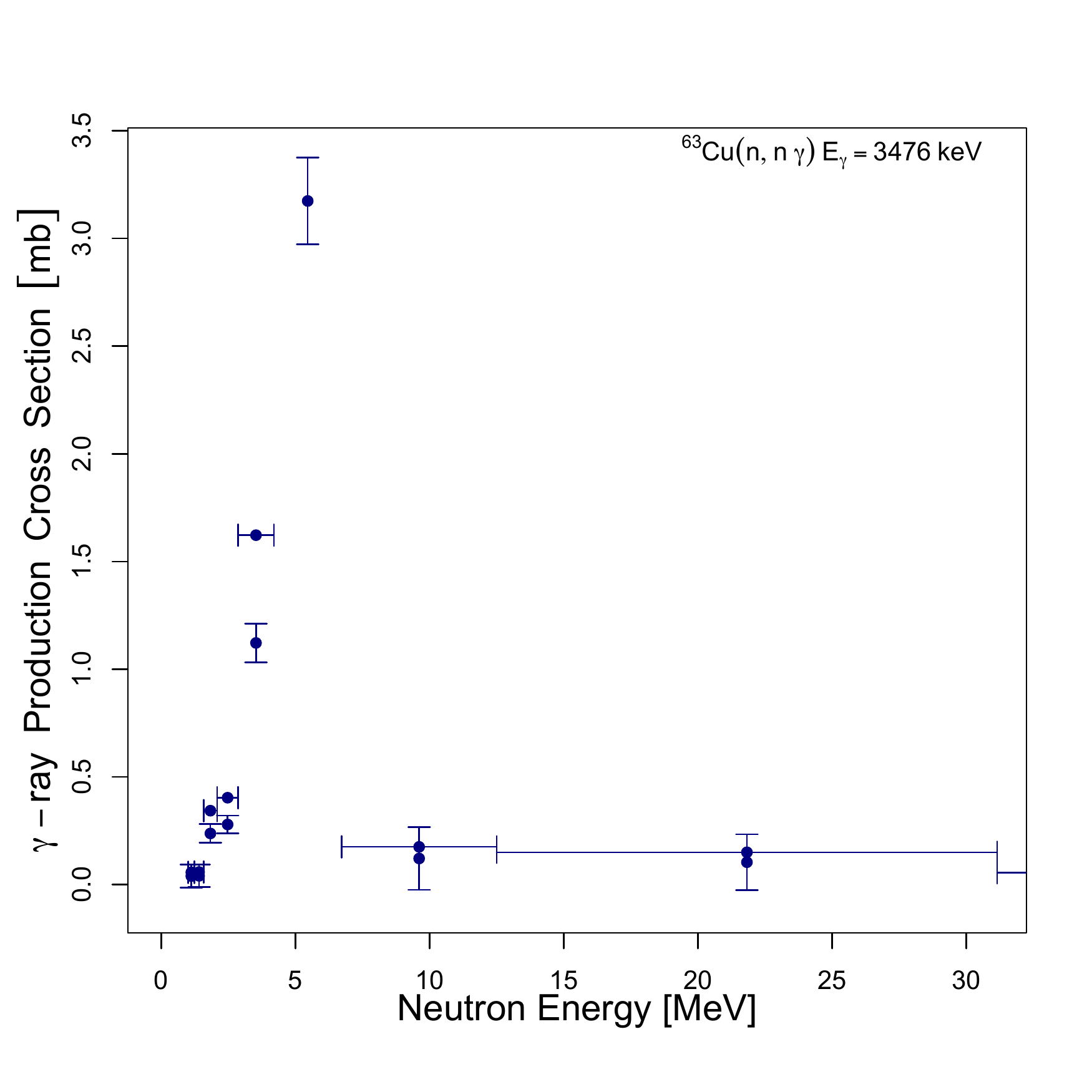}
\caption{The $\gamma$ production cross section for 3476-keV \gray in \tcu.  This $\gamma$ ray transition is in the vicinity of the $^{136}$Xe double-escape peak energy region, and is a ground state transition from the 3476-keV energy level in \tcu. The corresponding level for this \gray is not included in the ENDF/B-VII evaluation for \tcu}
\label{fig:ana:xe2}
\end{center}
\end{figure}

\paragraph{$^{136}$Xe Endpoint Energy}
	Recently the EXO collaboration reported the first ever observation of 2$\nu\beta\beta$ in  $^{136}$Xe\cite{exo2bb}.   This measurement found that the endpoint of the decay to be at 2457.83(37) keV.    There are no known \gray transitions from excited levels in the immediate vicinity of the  $^{136}$Xe endpoint energy.  The 3429-keV level has a 2468-keV $\gamma$ ray which is 11 keV away from the endpoint of $^{136}$Xe.  This level primarily decays to the ground state, and 23\% of the time decays to the second-excited level \cite{endf}.  The level cross section for the 3429-keV level in \tcu is shown if Fig. \ref{fig:ana:xe}.  Expanding the search window to include a 1.6\% energy resolution reveals an additional 3 levels with interfering $\gamma$-ray transitions.  There are no interfering $\gamma$-ray lines within 10 keV of the energy of the single-escape peak.   There are two $\gamma$-ray lines that are in 10-keV energy window corresponding to the energy of double escape peak.   The  $\gamma$ production cross section for the 3476-keV level in \tcu is shown in Fig. \ref{fig:ana:xe2}.   This level only decays to the ground state, and is not included in ENDF/B-VII evaluation.

\paragraph{$^{150}$Nd Endpoint Energy}	
	There are a variety of experiments that are studying neutrinoless double beta decay in $^{150}$Nd.  The current experimental techniques to study this isotope rely on kTons of liquid scintillator, and estimate a 3\% energy resolution \cite{snoplus}.  Our analysis did not reveal any known levels with \gray transitions in the immediate vicinity of the endpoint energy of this decay.    Table \ref{ana:tab:fep} gives the sensitivity of a 4-keV region around the 3367-keV, or the endpoint for this decay.  If there was a $\gamma$ ray in this energy region, its maximum $\gamma$ production cross section would be below 0.30 mb. Expanding the search region to 3\% energy resolution reveals an additional 23 levels in $^{nat}$Cu that produce full-energy $\gamma$ rays in the Nd region of interest.    Again, as in the case of calcium, all of these levels are above the cutoff energy for the ENDF/B-VII evaluated discrete levels in either Cu nucleus.

\section{Discussion and Conclusions}
	
	In conclusion, we have measured the $\gamma$ production cross sections for 111 transitions in $^{nat}$Cu and eight additional transitions in $^{62,64}$Cu for neutron energies 1 MeV $<$ $E_n$ $<$ 100 MeV.  We have compared our results with the ENDF/B-VII evaluations for Cu, and we found that the ENDF/B-VII evaluation agrees with our level cross sections for excited levels below 2 MeV for \tcu, and overestimates these level cross sections in \fcu.  For energy levels above 2 MeV we found significant discrepancies between the suggested level cross sections for both nuclei and our data.  Our data were also compared to both the TALYS-1.2, and CoH3 statistical model calculations.  Both calculations displayed considerable variability in the agreement with the experimental data; achieving excellent agreement for some levels, while differing by 300\% for other excited levels.   Given this degree of uncertainty in the models for most of the levels relevant for low-background, high-sensitivity studies need to be conducted to measure the level cross sections directly.    
	
	In addition to examining discrete levels, we also compared the total neutron inelastic cross sections for both Cu nuclei with the ENDF/B-VII evaluations.  We found excellent agreement for the \tcu total neutron inelastic  cross section with the ENDF/B-VII evaluation and previous measurements.  For the \fcu total neutron  inelastic cross section, our measurements were 30\% lower than the  ENDF/B-VII evaluation, which we attribute to unobserved transitions in \fcu.

	In addition to comparing our measurements with the ENDF/B-VII evaluation for Cu, we also looked at the implementation of these evaluations in the Monte Carlo program, Geant4.  We found that the Geant4 was properly reproducing the overall shape and strength ENDF/B-VII evaluated cross section.   On a more detailed level, however we found that the decay properties of the nucleus were not being properly modeled to the degree necessary for estimating backgrounds in rare-event searches.  There are relatively few discrete levels in ENDF/B-VII, and above 2 MeV the overwhelming strength of the continuum begins to dominate the simulated inelastic interactions with the nucleus.  In the current implementation, the ENDF/B-VII continuum file does not provide a decay path, nor do the de-excitation $\gamma$ rays correspond to transitions in Cu.  This approach needs to be reassessed for rare-decay searches, where the concern is \gray transitions that  might interfere, or even replicate the signature of the experimental process.

	Finally, we examined the potential implications of our measurements on $0\nu\beta\beta$ measurements.  We found that of the nine frequently studied $0\nu\beta\beta$ isotopes, only four had Q-values below the cutoff for ENDF/B-VII evaluated discrete levels in either Cu nucleus.  We were able to identify 16 \gray transitions in Cu as potential backgrounds for $0\nu\beta\beta$ experiments.  We measured level cross sections for nine of the levels that produce these \gray lines, and were able to put limits on the $\gamma$ production cross sections for regions where no specific \gray transitions were identified.  We identified several transitions that are problematic for experiments intending to use their $0\nu\beta\beta$ isotope in a bolometer.

\begin{acknowledgments}
The authors are grateful to Dr. Tatsumi Koi and Dr. Jason Detwiler for their helpful discussion of physics models in GEANT4.
We gratefully acknowledge the support of the US Department
of Energy through the LANL/LDRD Program for part of this
work.  We also acknowledge the support of the U.S. Department of Energy, Office of Nuclear Physics under
Contract No. 2011LANLE9BW. This work benefited from the use of the Los Alamos Neutron Science Center, funded
 by the U.S. Department of Energy under contract DE-AC52-06NA25396.
\end{acknowledgments}

\bibliographystyle{apsrev4-1}
\bibliography{cupaperv9}

\begin{thebibliography}{58}%
\makeatletter
\providecommand \@ifxundefined [1]{%
 \@ifx{#1\undefined}
}%
\providecommand \@ifnum [1]{%
 \ifnum #1\expandafter \@firstoftwo
 \else \expandafter \@secondoftwo
 \fi
}%
\providecommand \@ifx [1]{%
 \ifx #1\expandafter \@firstoftwo
 \else \expandafter \@secondoftwo
 \fi
}%
\providecommand \natexlab [1]{#1}%
\providecommand \enquote  [1]{``#1''}%
\providecommand \bibnamefont  [1]{#1}%
\providecommand \bibfnamefont [1]{#1}%
\providecommand \citenamefont [1]{#1}%
\providecommand \href@noop [0]{\@secondoftwo}%
\providecommand \href [0]{\begingroup \@sanitize@url \@href}%
\providecommand \@href[1]{\@@startlink{#1}\@@href}%
\providecommand \@@href[1]{\endgroup#1\@@endlink}%
\providecommand \@sanitize@url [0]{\catcode `\\12\catcode `\$12\catcode
  `\&12\catcode `\#12\catcode `\^12\catcode `\_12\catcode `\%12\relax}%
\providecommand \@@startlink[1]{}%
\providecommand \@@endlink[0]{}%
\providecommand \url  [0]{\begingroup\@sanitize@url \@url }%
\providecommand \@url [1]{\endgroup\@href {#1}{\urlprefix }}%
\providecommand \urlprefix  [0]{URL }%
\providecommand \Eprint [0]{\href }%
\@ifxundefined \urlstyle {%
  \providecommand \doi  [0]{\begingroup \@sanitize@url \@doi}%
  \providecommand \@doi [1]{\endgroup \@@startlink {\doibase
  #1}doi:\discretionary {}{}{}#1\@@endlink }%
}{%
  \providecommand \doi  [0]{doi:\discretionary{}{}{}\begingroup
  \urlstyle{rm}\Url }%
}%
\providecommand \doibase [0]{http://dx.doi.org/}%
\providecommand \Doi [0]{\begingroup \@sanitize@url \@Doi }%
\providecommand \@Doi  [1]{\endgroup\@@startlink{\doibase#1}\@@Doi}%
\providecommand \@@Doi [1]{#1\@@endlink}%
\providecommand \selectlanguage [0]{\@gobble}%
\providecommand \bibinfo  [0]{\@secondoftwo}%
\providecommand \bibfield  [0]{\@secondoftwo}%
\providecommand \translation [1]{[#1]}%
\providecommand \BibitemOpen [0]{}%
\providecommand \bibitemStop [0]{}%
\providecommand \bibitemNoStop [0]{.\EOS\space}%
\providecommand \EOS [0]{\spacefactor3000\relax}%
\providecommand \BibitemShut  [1]{\csname bibitem#1\endcsname}%
\bibitem [{\citenamefont {Elliott}\ and\ \citenamefont {Vogel}(2002)}]{Ell02}%
  \BibitemOpen
  \bibfield  {author} {\bibinfo {author} {\bibfnamefont {S.~R.}\ \bibnamefont
  {Elliott}}\ and\ \bibinfo {author} {\bibfnamefont {P.}~\bibnamefont
  {Vogel}},\ }\Doi {10.1146/annurev.nucl.52.050102.090641} {\bibfield
  {journal} {\bibinfo  {journal} {Annu. Rev. Nucl. Part. Sci.},\ }\textbf
  {\bibinfo {volume} {52}},\ \bibinfo {pages} {115} (\bibinfo {year}
  {2002})}\BibitemShut {NoStop}%
\bibitem [{\citenamefont {Elliott}\ and\ \citenamefont {Engel}(2004)}]{Ell04}%
  \BibitemOpen
  \bibfield  {author} {\bibinfo {author} {\bibfnamefont {S.~R.}\ \bibnamefont
  {Elliott}}\ and\ \bibinfo {author} {\bibfnamefont {J.}~\bibnamefont
  {Engel}},\ }\href {http://stacks.iop.org/0954-3899/30/i=9/a=R01} {\bibfield
  {journal} {\bibinfo  {journal} {J. Phys. G},\ }\textbf {\bibinfo {volume}
  {30}},\ \bibinfo {pages} {R183} (\bibinfo {year} {2004})}\BibitemShut
  {NoStop}%
\bibitem [{\citenamefont {Barabash}\ \emph {et~al.}(2004)\citenamefont
  {Barabash}, \citenamefont {Hubert}, \citenamefont {Huber},\ and\
  \citenamefont {Umatov}}]{Bar04}%
  \BibitemOpen
  \bibfield  {author} {\bibinfo {author} {\bibfnamefont {A.~S.}\ \bibnamefont
  {Barabash}}, \bibinfo {author} {\bibfnamefont {F.}~\bibnamefont {Hubert}},
  \bibinfo {author} {\bibfnamefont {P.}~\bibnamefont {Huber}}, \ and\ \bibinfo
  {author} {\bibfnamefont {V.~I.}\ \bibnamefont {Umatov}},\ }\href
  {http://dx.doi.org/10.1134/1.1675911} {\bibfield  {journal} {\bibinfo
  {journal} {JETP Letters},\ }\textbf {\bibinfo {volume} {79}},\ \bibinfo
  {pages} {10} (\bibinfo {year} {2004})}\BibitemShut {NoStop}%
\bibitem [{\citenamefont {Ejiri}(2005)}]{eji05}%
  \BibitemOpen
  \bibfield  {author} {\bibinfo {author} {\bibfnamefont {H.}~\bibnamefont
  {Ejiri}},\ }\href {http://jpsj.ipap.jp/link?JPSJ/74/2101/} {\bibfield
  {journal} {\bibinfo  {journal} {J. Phys. Soc. Jpn.},\ }\textbf {\bibinfo
  {volume} {74}},\ \bibinfo {pages} {2101} (\bibinfo {year}
  {2005})}\BibitemShut {NoStop}%
\bibitem [{\citenamefont {{\protect Avignone III}}\ \emph
  {et~al.}(2008)\citenamefont {{\protect Avignone III}}, \citenamefont
  {Elliott},\ and\ \citenamefont {Engel}}]{avi08}%
  \BibitemOpen
  \bibfield  {author} {\bibinfo {author} {\bibfnamefont {F.~T.}\ \bibnamefont
  {{\protect Avignone III}}}, \bibinfo {author} {\bibfnamefont {S.~R.}\
  \bibnamefont {Elliott}}, \ and\ \bibinfo {author} {\bibfnamefont
  {J.}~\bibnamefont {Engel}},\ }\href
  {http://link.aps.org/doi/10.1103/RevModPhys.80.481} {\bibfield  {journal}
  {\bibinfo  {journal} {Rev. Mod. Phys.},\ }\textbf {\bibinfo {volume} {80}},\
  \bibinfo {pages} {481} (\bibinfo {year} {2008})},\ \Eprint
  {http://arxiv.org/abs/arXiv:0708.1033} {arXiv:0708.1033} \BibitemShut
  {NoStop}%
\bibitem [{\citenamefont {Aalseth}\ \emph {et~al.}(2002)\citenamefont {Aalseth}
  \emph {et~al.}}]{aal02a}%
  \BibitemOpen
  \bibfield  {author} {\bibinfo {author} {\bibfnamefont {C.~E.}\ \bibnamefont
  {Aalseth}} \emph {et~al.} (\bibinfo {collaboration} {IGEX}),\ }\href
  {http://link.aps.org/doi/10.1103/PhysRevD.65.092007} {\bibfield  {journal}
  {\bibinfo  {journal} {Phys. Rev. D.},\ }\textbf {\bibinfo {volume} {65}},\
  \bibinfo {pages} {092007} (\bibinfo {year} {2002})}\BibitemShut {NoStop}%
\bibitem [{\citenamefont {Baudis}\ \emph {et~al.}(1999)\citenamefont {Baudis}
  \emph {et~al.}}]{bau99}%
  \BibitemOpen
  \bibfield  {author} {\bibinfo {author} {\bibfnamefont {L.}~\bibnamefont
  {Baudis}} \emph {et~al.},\ }\href
  {http://link.aps.org/doi/10.1103/PhysRevLett.83.41} {\bibfield  {journal}
  {\bibinfo  {journal} {Phys. Rev. Lett.},\ }\textbf {\bibinfo {volume} {83}},\
  \bibinfo {pages} {41} (\bibinfo {year} {1999})}\BibitemShut {NoStop}%
\bibitem [{\citenamefont {Klapdor-Kleingrothaus}\ and\ \citenamefont
  {Krivosheina}(2006)}]{kla06}%
  \BibitemOpen
  \bibfield  {author} {\bibinfo {author} {\bibfnamefont {H.~V.}\ \bibnamefont
  {Klapdor-Kleingrothaus}}\ and\ \bibinfo {author} {\bibfnamefont {I.~V.}\
  \bibnamefont {Krivosheina}},\ }\href
  {http://www.worldscientific.com/doi/abs/10.1142/S0217732306020937} {\bibfield
   {journal} {\bibinfo  {journal} {Mod. Phys. Lett. A},\ }\textbf {\bibinfo
  {volume} {20}},\ \bibinfo {pages} {1547} (\bibinfo {year}
  {2006})}\BibitemShut {NoStop}%
\bibitem [{\citenamefont {Elliott}\ \emph {et~al.}(2009)\citenamefont {Elliott}
  \emph {et~al.}}]{ell08}%
  \BibitemOpen
  \bibfield  {author} {\bibinfo {author} {\bibfnamefont {S.}~\bibnamefont
  {Elliott}} \emph {et~al.} (\bibinfo {collaboration} {{\majorana
  Collaboration}}),\ }\href {http://stacks.iop.org/1742-6596/173/i=1/a=012007}
  {\bibfield  {journal} {\bibinfo  {journal} {J. Phys. Conf. Proc.},\ }\textbf
  {\bibinfo {volume} {173}},\ \bibinfo {pages} {012007} (\bibinfo {year}
  {2009})},\ \Eprint {http://arxiv.org/abs/arXiv:0807.1741} {arXiv:0807.1741}
  \BibitemShut {NoStop}%
\bibitem [{\citenamefont {Schonert}\ \emph {et~al.}(2005)\citenamefont
  {Schonert} \emph {et~al.}}]{Gerda}%
  \BibitemOpen
  \bibfield  {author} {\bibinfo {author} {\bibfnamefont {S.}~\bibnamefont
  {Schonert}} \emph {et~al.} (\bibinfo {collaboration} {Gerda Collaboration}),\
  }\href {http://www.sciencedirect.com/science/article/pii/S0920563205005116}
  {\bibfield  {journal} {\bibinfo  {journal} {Nucl. Phys. B, Proc. Suppl.},\
  }\textbf {\bibinfo {volume} {145}},\ \bibinfo {pages} {242 } (\bibinfo {year}
  {2005})}\BibitemShut {NoStop}%
\bibitem [{\citenamefont {Fogli}\ \emph {et~al.}(2011)\citenamefont {Fogli},
  \citenamefont {Lisi}, \citenamefont {Marrone}, \citenamefont {Palazzo},\ and\
  \citenamefont {Rotunno}}]{fog11}%
  \BibitemOpen
  \bibfield  {author} {\bibinfo {author} {\bibfnamefont {G.~L.}\ \bibnamefont
  {Fogli}}, \bibinfo {author} {\bibfnamefont {E.}~\bibnamefont {Lisi}},
  \bibinfo {author} {\bibfnamefont {A.}~\bibnamefont {Marrone}}, \bibinfo
  {author} {\bibfnamefont {A.}~\bibnamefont {Palazzo}}, \ and\ \bibinfo
  {author} {\bibfnamefont {A.~M.}\ \bibnamefont {Rotunno}},\ }\href
  {http://link.aps.org/doi/10.1103/PhysRevD.84.053007} {\bibfield  {journal}
  {\bibinfo  {journal} {Phys. Rev. D},\ }\textbf {\bibinfo {volume} {84}},\
  \bibinfo {pages} {053007} (\bibinfo {year} {2011})}\BibitemShut {NoStop}%
\bibitem [{\citenamefont {Mei}\ and\ \citenamefont {Hime}(2006)}]{mei06}%
  \BibitemOpen
  \bibfield  {author} {\bibinfo {author} {\bibfnamefont {D.-M.}\ \bibnamefont
  {Mei}}\ and\ \bibinfo {author} {\bibfnamefont {A.}~\bibnamefont {Hime}},\
  }\href {http://link.aps.org/doi/10.1103/PhysRevD.73.053004} {\bibfield
  {journal} {\bibinfo  {journal} {Phys. Rev. D},\ }\textbf {\bibinfo {volume}
  {73}},\ \bibinfo {pages} {053004} (\bibinfo {year} {2006})}\BibitemShut
  {NoStop}%
\bibitem [{\citenamefont {Mei}\ \emph {et~al.}(2008)\citenamefont {Mei},
  \citenamefont {Elliott}, \citenamefont {Hime}, \citenamefont {Gehman},\ and\
  \citenamefont {Kazkaz}}]{mei08}%
  \BibitemOpen
  \bibfield  {author} {\bibinfo {author} {\bibfnamefont {D.-M.}\ \bibnamefont
  {Mei}}, \bibinfo {author} {\bibfnamefont {S.}~\bibnamefont {Elliott}},
  \bibinfo {author} {\bibfnamefont {A.}~\bibnamefont {Hime}}, \bibinfo {author}
  {\bibfnamefont {V.}~\bibnamefont {Gehman}}, \ and\ \bibinfo {author}
  {\bibfnamefont {K.}~\bibnamefont {Kazkaz}},\ }\href
  {http://link.aps.org/doi/10.1103/PhysRevC.77.054614} {\bibfield  {journal}
  {\bibinfo  {journal} {Phys. Rev. C},\ }\textbf {\bibinfo {volume} {77}},\
  \bibinfo {pages} {054614} (\bibinfo {year} {2008})},\ \Eprint
  {http://arxiv.org/abs/nucl-ex/0704.0306} {arXiv:nucl-ex/0704.0306}
  \BibitemShut {NoStop}%
\bibitem [{\citenamefont {Bonner}\ and\ \citenamefont
  {Slattery}(1959)}]{bon59}%
  \BibitemOpen
  \bibfield  {author} {\bibinfo {author} {\bibfnamefont {T.~W.}\ \bibnamefont
  {Bonner}}\ and\ \bibinfo {author} {\bibfnamefont {J.~C.}\ \bibnamefont
  {Slattery}},\ }\href {http://link.aps.org/doi/10.1103/PhysRev.113.1088}
  {\bibfield  {journal} {\bibinfo  {journal} {Phys. Rev.},\ }\textbf {\bibinfo
  {volume} {113}},\ \bibinfo {pages} {1088} (\bibinfo {year}
  {1959})}\BibitemShut {NoStop}%
\bibitem [{\citenamefont {Beyster}\ \emph {et~al.}(1956)\citenamefont
  {Beyster}, \citenamefont {Walt},\ and\ \citenamefont {Salmi}}]{bey56}%
  \BibitemOpen
  \bibfield  {author} {\bibinfo {author} {\bibfnamefont {J.~R.}\ \bibnamefont
  {Beyster}}, \bibinfo {author} {\bibfnamefont {M.}~\bibnamefont {Walt}}, \
  and\ \bibinfo {author} {\bibfnamefont {E.~W.}\ \bibnamefont {Salmi}},\ }\href
  {http://link.aps.org/doi/10.1103/PhysRev.104.1319} {\bibfield  {journal}
  {\bibinfo  {journal} {Phys. Rev.},\ }\textbf {\bibinfo {volume} {104}},\
  \bibinfo {pages} {1319} (\bibinfo {year} {1956})}\BibitemShut {NoStop}%
\bibitem [{\citenamefont {{Daniels}}\ and\ \citenamefont
  {{Felsteiner}}(1968)}]{1968DA}%
  \BibitemOpen
  \bibfield  {author} {\bibinfo {author} {\bibfnamefont {J.~M.}\ \bibnamefont
  {{Daniels}}}\ and\ \bibinfo {author} {\bibfnamefont {J.}~\bibnamefont
  {{Felsteiner}}},\ }\href
  {http://www.nrcresearchpress.com/doi/abs/10.1139/p68-525} {\bibfield
  {journal} {\bibinfo  {journal} {Can. J. Phys.},\ }\textbf {\bibinfo {volume}
  {46}},\ \bibinfo {pages} {1849} (\bibinfo {year} {1968})}\BibitemShut
  {NoStop}%
\bibitem [{\citenamefont {Felsteiner}\ and\ \citenamefont
  {Serfaty}(1970)}]{1970Fe}%
  \BibitemOpen
  \bibfield  {author} {\bibinfo {author} {\bibfnamefont {J.}~\bibnamefont
  {Felsteiner}}\ and\ \bibinfo {author} {\bibfnamefont {R.}~\bibnamefont
  {Serfaty}},\ }\href
  {http://www.sciencedirect.com/science/article/pii/0375947470906354}
  {\bibfield  {journal} {\bibinfo  {journal} {Nucl. Phys.},\ }\textbf {\bibinfo
  {volume} {A148}},\ \bibinfo {pages} {428 } (\bibinfo {year}
  {1970})}\BibitemShut {NoStop}%
\bibitem [{\citenamefont {{French}}\ \emph {et~al.}(1971)\citenamefont
  {{French}}, \citenamefont {{Wehrbein}}, \citenamefont {{Goss}},\ and\
  \citenamefont {{Gasper}}}]{1971Fr}%
  \BibitemOpen
  \bibfield  {author} {\bibinfo {author} {\bibfnamefont {W.~R.}\ \bibnamefont
  {{French}}}, \bibinfo {author} {\bibfnamefont {W.~M.}\ \bibnamefont
  {{Wehrbein}}}, \bibinfo {author} {\bibfnamefont {D.}~\bibnamefont {{Goss}}},
  \ and\ \bibinfo {author} {\bibfnamefont {J.~K.}\ \bibnamefont {{Gasper}}},\
  }\href {http://ajp.aapt.org/resource/1/ajpias/v39/i2/p131_s1} {\bibfield
  {journal} {\bibinfo  {journal} {Am. J. Phys.},\ }\textbf {\bibinfo {volume}
  {39}},\ \bibinfo {pages} {131} (\bibinfo {year} {1971})}\BibitemShut
  {NoStop}%
\bibitem [{\citenamefont {Kramarovsky}\ \emph {et~al.}(1989)\citenamefont
  {Kramarovsky}, \citenamefont {Nemilov}, \citenamefont {Pobedonostsev},\ and\
  \citenamefont {Shiryaev}}]{kram89}%
  \BibitemOpen
  \bibfield  {author} {\bibinfo {author} {\bibfnamefont {Y.~M.}\ \bibnamefont
  {Kramarovsky}}, \bibinfo {author} {\bibfnamefont {Y.~A.}\ \bibnamefont
  {Nemilov}}, \bibinfo {author} {\bibfnamefont {L.~A.}\ \bibnamefont
  {Pobedonostsev}}, \ and\ \bibinfo {author} {\bibfnamefont {B.~M.}\
  \bibnamefont {Shiryaev}},\ }\href@noop {} {\bibfield  {journal} {\bibinfo
  {journal} {Yad. Fiz. [Sov. J. Nucl. Phys.]},\ }\textbf {\bibinfo {volume}
  {22}},\ \bibinfo {pages} {8912} (\bibinfo {year} {1989})},\ \bibinfo {note}
  {data taken from the EXFOR database, file EXFOR 41049.002-41049.011 dated
  Sept 26, 2005, retrieved from the IAEA Nuclear Data Services
  Website}\BibitemShut {NoStop}%
\bibitem [{\citenamefont {Zhou}\ \emph {et~al.}(2007)\citenamefont {Zhou},
  \citenamefont {Deng}, \citenamefont {Ding}, \citenamefont {Hua},
  \citenamefont {Zhu}, \citenamefont {Wang}, \citenamefont {Zhao},\ and\
  \citenamefont {Fan}}]{zhou}%
  \BibitemOpen
  \bibfield  {author} {\bibinfo {author} {\bibfnamefont {H.}~\bibnamefont
  {Zhou}}, \bibinfo {author} {\bibfnamefont {F.}~\bibnamefont {Deng}}, \bibinfo
  {author} {\bibfnamefont {X.}~\bibnamefont {Ding}}, \bibinfo {author}
  {\bibfnamefont {M.}~\bibnamefont {Hua}}, \bibinfo {author} {\bibfnamefont
  {Q.}~\bibnamefont {Zhu}}, \bibinfo {author} {\bibfnamefont {C.}~\bibnamefont
  {Wang}}, \bibinfo {author} {\bibfnamefont {Q.}~\bibnamefont {Zhao}}, \ and\
  \bibinfo {author} {\bibfnamefont {G.}~\bibnamefont {Fan}},\ }\href
  {http://www.new.ans.org/pubs/journals/nse/a_2733} {\bibfield  {journal}
  {\bibinfo  {journal} {Nucl. Sci. Eng.},\ }\textbf {\bibinfo {volume} {157}},\
  \bibinfo {pages} {354} (\bibinfo {year} {2007})}\BibitemShut {NoStop}%
\bibitem [{\citenamefont {Almen-Ramstrom}(1975)}]{almen}%
  \BibitemOpen
  \bibfield  {author} {\bibinfo {author} {\bibfnamefont {E.}~\bibnamefont
  {Almen-Ramstrom}},\ }\href@noop {} {}\bibinfo {type} {Tech. Rep.}\ \bibinfo
  {number} {503}\ (\bibinfo  {institution} {Aktiebolaget Atomenergi},\ \bibinfo
  {address} {Stockholm/Studsvik},\ \bibinfo {year} {1975})\ \bibinfo {note}
  {data taken from the EXFOR database, files EXFOR 20788.011 and 20788.012
  dated Sept 26, 2005, retrieved from the IAEA Nuclear Data Services
  Website}\BibitemShut {NoStop}%
\bibitem [{\citenamefont {Nishimura}\ \emph {et~al.}(1965)\citenamefont
  {Nishimura}, \citenamefont {Okano},\ and\ \citenamefont
  {Kikuchi}}]{nishimura}%
  \BibitemOpen
  \bibfield  {author} {\bibinfo {author} {\bibfnamefont {K.}~\bibnamefont
  {Nishimura}}, \bibinfo {author} {\bibfnamefont {K.}~\bibnamefont {Okano}}, \
  and\ \bibinfo {author} {\bibfnamefont {S.}~\bibnamefont {Kikuchi}},\ }\href
  {http://www.sciencedirect.com/science/article/pii/0029558265905389}
  {\bibfield  {journal} {\bibinfo  {journal} {Nucl. Phys.},\ }\textbf {\bibinfo
  {volume} {70}},\ \bibinfo {pages} {421 } (\bibinfo {year}
  {1965})}\BibitemShut {NoStop}%
\bibitem [{\citenamefont {Tucker}\ \emph {et~al.}(1965)\citenamefont {Tucker},
  \citenamefont {Wells},\ and\ \citenamefont {Meyerhof}}]{tucker}%
  \BibitemOpen
  \bibfield  {author} {\bibinfo {author} {\bibfnamefont {A.~B.}\ \bibnamefont
  {Tucker}}, \bibinfo {author} {\bibfnamefont {J.~T.}\ \bibnamefont {Wells}}, \
  and\ \bibinfo {author} {\bibfnamefont {W.~E.}\ \bibnamefont {Meyerhof}},\
  }\href {http://link.aps.org/doi/10.1103/PhysRev.137.B1181} {\bibfield
  {journal} {\bibinfo  {journal} {Phys. Rev.},\ }\textbf {\bibinfo {volume}
  {137}},\ \bibinfo {pages} {B1181} (\bibinfo {year} {1965})},\ \bibinfo {note}
  {data taken from the EXFOR database, file EXFOR 11720.004 dated April 27,
  2005, retrieved from the IAEA Nuclear Data Services Website}\BibitemShut
  {NoStop}%
\bibitem [{\citenamefont {Jonsson}\ \emph {et~al.}(1969)\citenamefont
  {Jonsson}, \citenamefont {Nyberg},\ and\ \citenamefont
  {Bergqvist}}]{joensson}%
  \BibitemOpen
  \bibfield  {author} {\bibinfo {author} {\bibfnamefont {B.}~\bibnamefont
  {Jonsson}}, \bibinfo {author} {\bibfnamefont {K.}~\bibnamefont {Nyberg}}, \
  and\ \bibinfo {author} {\bibfnamefont {I.}~\bibnamefont {Bergqvist}},\
  }\href@noop {} {\bibfield  {journal} {\bibinfo  {journal} {Ark. Fys.},\
  }\textbf {\bibinfo {volume} {39}},\ \bibinfo {pages} {295} (\bibinfo {year}
  {1969})},\ \bibinfo {note} {data taken from the EXFOR database, files EXFOR
  20164.021 and 20164.025 dated Feb. 11, 2008, retrieved from the IAEA Nuclear
  Data Services Website}\BibitemShut {NoStop}%
\bibitem [{\citenamefont {Xiamin}\ \emph {et~al.}()\citenamefont {Xiamin},
  \citenamefont {Ronglin}, \citenamefont {Jinqiang}, \citenamefont {Yongshun},\
  and\ \citenamefont {Dazhao}}]{xiamin}%
  \BibitemOpen
  \bibfield  {author} {\bibinfo {author} {\bibfnamefont {S.}~\bibnamefont
  {Xiamin}}, \bibinfo {author} {\bibfnamefont {S.}~\bibnamefont {Ronglin}},
  \bibinfo {author} {\bibfnamefont {X.}~\bibnamefont {Jinqiang}}, \bibinfo
  {author} {\bibfnamefont {W.}~\bibnamefont {Yongshun}}, \ and\ \bibinfo
  {author} {\bibfnamefont {D.}~\bibnamefont {Dazhao}},\ }\href@noop {}
  {\bibfield  {journal} {\bibinfo  {journal} {Chin. J. Nucl. Phys.},\ }\textbf
  {\bibinfo {volume} {4}}},\ \bibinfo {note} {data taken from the EXFOR
  database, file EXFOR 30656.024 and 30656.025 dated JAN 13, 2012, retrieved
  from the IAEA Nuclear Data Services Website}\BibitemShut {NoStop}%
\bibitem [{\citenamefont {Guenther}\ \emph {et~al.}(1986)\citenamefont
  {Guenther}, \citenamefont {Smith}, \citenamefont {Smith},\ and\ \citenamefont
  {Whalen}}]{guenther}%
  \BibitemOpen
  \bibfield  {author} {\bibinfo {author} {\bibfnamefont {P.~T.}\ \bibnamefont
  {Guenther}}, \bibinfo {author} {\bibfnamefont {D.~L.}\ \bibnamefont {Smith}},
  \bibinfo {author} {\bibfnamefont {A.~B.}\ \bibnamefont {Smith}}, \ and\
  \bibinfo {author} {\bibfnamefont {J.~F.}\ \bibnamefont {Whalen}},\ }\href
  {http://www.sciencedirect.com/science/article/pii/0375947486900928}
  {\bibfield  {journal} {\bibinfo  {journal} {Nucl. Phys.},\ }\textbf {\bibinfo
  {volume} {A448}},\ \bibinfo {pages} {280} (\bibinfo {year} {1986})},\
  \bibinfo {note} {data taken from the EXFOR database, files EXFOR 12869.007,
  12869.008, and 12869.009 dated Jan 19, 2010, retrieved from the IAEA Nuclear
  Data Services Website}\BibitemShut {NoStop}%
\bibitem [{\citenamefont {{Delaroche}}\ \emph {et~al.}(1982)\citenamefont
  {{Delaroche}}, \citenamefont {{El-Kadi}}, \citenamefont {{Guss}},
  \citenamefont {{Floyd}},\ and\ \citenamefont {{Walter}}}]{1982De}%
  \BibitemOpen
  \bibfield  {author} {\bibinfo {author} {\bibfnamefont {J.~P.}\ \bibnamefont
  {{Delaroche}}}, \bibinfo {author} {\bibfnamefont {S.~M.}\ \bibnamefont
  {{El-Kadi}}}, \bibinfo {author} {\bibfnamefont {P.~P.}\ \bibnamefont
  {{Guss}}}, \bibinfo {author} {\bibfnamefont {C.~E.}\ \bibnamefont {{Floyd}}},
  \ and\ \bibinfo {author} {\bibfnamefont {R.~L.}\ \bibnamefont {{Walter}}},\
  }\href {http://www.sciencedirect.com/science/article/pii/0375947482902822}
  {\bibfield  {journal} {\bibinfo  {journal} {Nucl. Phys.},\ }\textbf {\bibinfo
  {volume} {A390}},\ \bibinfo {pages} {541} (\bibinfo {year}
  {1982})}\BibitemShut {NoStop}%
\bibitem [{\citenamefont {Dickens}(1983)}]{1983Di}%
  \BibitemOpen
  \bibfield  {author} {\bibinfo {author} {\bibfnamefont {J.~K.}\ \bibnamefont
  {Dickens}},\ }\href
  {http://www.sciencedirect.com/science/article/pii/0375947483905262}
  {\bibfield  {journal} {\bibinfo  {journal} {Nucl. Phys.},\ }\textbf {\bibinfo
  {volume} {A401}},\ \bibinfo {pages} {189 } (\bibinfo {year}
  {1983})}\BibitemShut {NoStop}%
\bibitem [{\citenamefont {{Dostemesova}}\ \emph {et~al.}(1987)\citenamefont
  {{Dostemesova}}, \citenamefont {{Kaipov}},\ and\ \citenamefont
  {{Kosyak}}}]{1987Do}%
  \BibitemOpen
  \bibfield  {author} {\bibinfo {author} {\bibfnamefont {G.~A.}\ \bibnamefont
  {{Dostemesova}}}, \bibinfo {author} {\bibfnamefont {D.~K.}\ \bibnamefont
  {{Kaipov}}}, \ and\ \bibinfo {author} {\bibfnamefont {Y.~G.}\ \bibnamefont
  {{Kosyak}}},\ }in\ \href@noop {} {\emph {\bibinfo {booktitle} {Program and
  Theses}}}\ (\bibinfo {year} {1987})\BibitemShut {NoStop}%
\bibitem [{\citenamefont {{El-Kadi}}\ \emph {et~al.}(1982)\citenamefont
  {{El-Kadi}}, \citenamefont {{Nelson}}, \citenamefont {{Purser}},
  \citenamefont {{Walter}}, \citenamefont {{Beyerle}}, \citenamefont
  {{Gould}},\ and\ \citenamefont {{Seagondollar}}}]{1982el}%
  \BibitemOpen
  \bibfield  {author} {\bibinfo {author} {\bibfnamefont {S.~M.}\ \bibnamefont
  {{El-Kadi}}}, \bibinfo {author} {\bibfnamefont {C.~E.}\ \bibnamefont
  {{Nelson}}}, \bibinfo {author} {\bibfnamefont {F.~O.}\ \bibnamefont
  {{Purser}}}, \bibinfo {author} {\bibfnamefont {R.~L.}\ \bibnamefont
  {{Walter}}}, \bibinfo {author} {\bibfnamefont {A.}~\bibnamefont {{Beyerle}}},
  \bibinfo {author} {\bibfnamefont {C.~R.}\ \bibnamefont {{Gould}}}, \ and\
  \bibinfo {author} {\bibfnamefont {L.~W.}\ \bibnamefont {{Seagondollar}}},\
  }\href {http://www.sciencedirect.com/science/article/pii/0375947482902810}
  {\bibfield  {journal} {\bibinfo  {journal} {Nucl. Phys.},\ }\textbf {\bibinfo
  {volume} {A390}},\ \bibinfo {pages} {509} (\bibinfo {year}
  {1982})}\BibitemShut {NoStop}%
\bibitem [{\citenamefont {{Kosyak}}\ \emph {et~al.}(2001)\citenamefont
  {{Kosyak}}, \citenamefont {{Chekushina}}, \citenamefont {{Adymov}},\ and\
  \citenamefont {{Ermatov}}}]{kos01}%
  \BibitemOpen
  \bibfield  {author} {\bibinfo {author} {\bibfnamefont {Y.~G.}\ \bibnamefont
  {{Kosyak}}}, \bibinfo {author} {\bibfnamefont {L.~V.}\ \bibnamefont
  {{Chekushina}}}, \bibinfo {author} {\bibfnamefont {Z.~I.}\ \bibnamefont
  {{Adymov}}}, \ and\ \bibinfo {author} {\bibfnamefont {A.~S.}\ \bibnamefont
  {{Ermatov}}},\ }\href@noop {} {\bibfield  {journal} {\bibinfo  {journal}
  {Bull. Rus. Acad. Sci. Phys.},\ }\textbf {\bibinfo {volume} {65}},\ \bibinfo
  {pages} {128} (\bibinfo {year} {2001})}\BibitemShut {NoStop}%
\bibitem [{\citenamefont {{Kosyak}}\ \emph
  {et~al.}(2000){\natexlab{a}}\citenamefont {{Kosyak}}, \citenamefont
  {{Chekushina}},\ and\ \citenamefont {{Ermatov}}}]{kosyak00}%
  \BibitemOpen
  \bibfield  {author} {\bibinfo {author} {\bibfnamefont {Y.~G.}\ \bibnamefont
  {{Kosyak}}}, \bibinfo {author} {\bibfnamefont {L.~V.}\ \bibnamefont
  {{Chekushina}}}, \ and\ \bibinfo {author} {\bibfnamefont {A.~S.}\
  \bibnamefont {{Ermatov}}},\ }\href@noop {} {\bibfield  {journal} {\bibinfo
  {journal} {Bull. Rus. Acad. Sci. Phys.},\ }\textbf {\bibinfo {volume} {64}},\
  \bibinfo {pages} {321} (\bibinfo {year} {2000}{\natexlab{a}})}\BibitemShut
  {NoStop}%
\bibitem [{\citenamefont {{\protect Avignone III}}(2008)}]{avi07}%
  \BibitemOpen
  \bibfield  {author} {\bibinfo {author} {\bibfnamefont {F.~T.}\ \bibnamefont
  {{\protect Avignone III}}},\ }\href
  {http://stacks.iop.org/1742-6596/120/i=5/a=052059} {\bibfield  {journal}
  {\bibinfo  {journal} {J. Phys. Conf. Proc.},\ }\textbf {\bibinfo {volume}
  {120}},\ \bibinfo {pages} {052059} (\bibinfo {year} {2008})},\ \Eprint
  {http://arxiv.org/abs/arXiv: 0711.4808v1} {arXiv: 0711.4808v1} \BibitemShut
  {NoStop}%
\bibitem [{\citenamefont {Lisowski}\ \emph {et~al.}(1990)\citenamefont
  {Lisowski}, \citenamefont {Bowman}, \citenamefont {Wender},\ and\
  \citenamefont {Russell}}]{1990Lis}%
  \BibitemOpen
  \bibfield  {author} {\bibinfo {author} {\bibfnamefont {P.~W.}\ \bibnamefont
  {Lisowski}}, \bibinfo {author} {\bibfnamefont {C.~D.}\ \bibnamefont
  {Bowman}}, \bibinfo {author} {\bibfnamefont {S.~A.}\ \bibnamefont {Wender}},
  \ and\ \bibinfo {author} {\bibfnamefont {G.~J.}\ \bibnamefont {Russell}},\
  }\href@noop {} {\bibfield  {journal} {\bibinfo  {journal} {Nucl. Sci. Eng.},\
  }\textbf {\bibinfo {volume} {106}},\ \bibinfo {pages} {208} (\bibinfo {year}
  {1990})}\BibitemShut {NoStop}%
\bibitem [{\citenamefont {Becker}\ and\ \citenamefont {Nelson}(2005)}]{2005Be}%
  \BibitemOpen
  \bibfield  {author} {\bibinfo {author} {\bibfnamefont {J.~A.}\ \bibnamefont
  {Becker}}\ and\ \bibinfo {author} {\bibfnamefont {R.~O.}\ \bibnamefont
  {Nelson}},\ }in\ \href@noop {} {\emph {\bibinfo {booktitle} {International
  Conference on Nuclear Data for Science and Technology}}},\ \bibinfo {editor}
  {edited by\ \bibinfo {editor} {\bibfnamefont {R.~C.}\ \bibnamefont {Haight}},
  \bibinfo {editor} {\bibfnamefont {M.~B.}\ \bibnamefont {Chadwick}}, \bibinfo
  {editor} {\bibfnamefont {T.}~\bibnamefont {Kawano}}, \ and\ \bibinfo {editor}
  {\bibfnamefont {P.}~\bibnamefont {Talou}}}\ (\bibinfo  {publisher} {American
  Institute of Physics},\ \bibinfo {year} {2005})\ pp.\ \bibinfo {pages}
  {748--753}\BibitemShut {NoStop}%
\bibitem [{\citenamefont {Guiseppe}\ \emph {et~al.}(2009)\citenamefont
  {Guiseppe}, \citenamefont {Devlin}, \citenamefont {Elliott}, \citenamefont
  {Fotiades}, \citenamefont {Hime}, \citenamefont {Mei}, \citenamefont
  {Nelson},\ and\ \citenamefont {Perepelitsa}}]{2009Gui}%
  \BibitemOpen
  \bibfield  {author} {\bibinfo {author} {\bibfnamefont {V.~E.}\ \bibnamefont
  {Guiseppe}}, \bibinfo {author} {\bibfnamefont {M.}~\bibnamefont {Devlin}},
  \bibinfo {author} {\bibfnamefont {S.~R.}\ \bibnamefont {Elliott}}, \bibinfo
  {author} {\bibfnamefont {N.}~\bibnamefont {Fotiades}}, \bibinfo {author}
  {\bibfnamefont {A.}~\bibnamefont {Hime}}, \bibinfo {author} {\bibfnamefont
  {D.-M.}\ \bibnamefont {Mei}}, \bibinfo {author} {\bibfnamefont {R.~O.}\
  \bibnamefont {Nelson}}, \ and\ \bibinfo {author} {\bibfnamefont {D.~V.}\
  \bibnamefont {Perepelitsa}},\ }\href
  {http://link.aps.org/doi/10.1103/PhysRevC.79.054604} {\bibfield  {journal}
  {\bibinfo  {journal} {Phys. Rev. C},\ }\textbf {\bibinfo {volume} {79}},\
  \bibinfo {pages} {054604} (\bibinfo {year} {2009})}\BibitemShut {NoStop}%
\bibitem [{\citenamefont {Fotiades}\ \emph {et~al.}(2004)\citenamefont
  {Fotiades} \emph {et~al.}}]{fot04}%
  \BibitemOpen
  \bibfield  {author} {\bibinfo {author} {\bibfnamefont {N.}~\bibnamefont
  {Fotiades}} \emph {et~al.},\ }\href
  {http://link.aps.org/doi/10.1103/PhysRevC.69.024601} {\bibfield  {journal}
  {\bibinfo  {journal} {Phys. Rev. C},\ }\textbf {\bibinfo {volume} {69}},\
  \bibinfo {pages} {024601} (\bibinfo {year} {2004})},\ \Eprint
  {http://arxiv.org/abs/nucle-ex/0809.5074} {nucle-ex/0809.5074} \BibitemShut
  {NoStop}%
\bibitem [{\citenamefont {{Koning}}\ \emph {et~al.}(2005)\citenamefont
  {{Koning}}, \citenamefont {{Hilaire}},\ and\ \citenamefont
  {{Duijvestijn}}}]{talys}%
  \BibitemOpen
  \bibfield  {author} {\bibinfo {author} {\bibfnamefont {A.~J.}\ \bibnamefont
  {{Koning}}}, \bibinfo {author} {\bibfnamefont {S.}~\bibnamefont {{Hilaire}}},
  \ and\ \bibinfo {author} {\bibfnamefont {M.~C.}\ \bibnamefont
  {{Duijvestijn}}},\ }in\ \href@noop {} {\emph {\bibinfo {booktitle} {Proc.
  Intern. Conf. Nuclear Data for Science and Technology}}}\ (\bibinfo {year}
  {2005})\BibitemShut {NoStop}%
\bibitem [{\citenamefont {Kawano}(2010)}]{kawano}%
  \BibitemOpen
  \bibfield  {author} {\bibinfo {author} {\bibfnamefont {T.}~\bibnamefont
  {Kawano}},\ }\href@noop {} {}\bibinfo {howpublished} {Private Communication}
  (\bibinfo {year} {2010})\BibitemShut {NoStop}%
\bibitem [{\citenamefont {Erjun}\ and\ \citenamefont
  {Junde}(2001)}]{erjun2001}%
  \BibitemOpen
  \bibfield  {author} {\bibinfo {author} {\bibfnamefont {B.}~\bibnamefont
  {Erjun}}\ and\ \bibinfo {author} {\bibfnamefont {H.}~\bibnamefont {Junde}},\
  }\href {http://www.sciencedirect.com/science/article/pii/S009037520190002X}
  {\bibfield  {journal} {\bibinfo  {journal} {Nucl. Data Sheets},\ }\textbf
  {\bibinfo {volume} {92}},\ \bibinfo {pages} {147 } (\bibinfo {year}
  {2001})}\BibitemShut {NoStop}%
\bibitem [{\citenamefont {Browne}\ and\ \citenamefont
  {Tuli}(2010)}]{browne2010}%
  \BibitemOpen
  \bibfield  {author} {\bibinfo {author} {\bibfnamefont {E.}~\bibnamefont
  {Browne}}\ and\ \bibinfo {author} {\bibfnamefont {J.}~\bibnamefont {Tuli}},\
  }\href {http://www.sciencedirect.com/science/article/pii/S0090375210000864}
  {\bibfield  {journal} {\bibinfo  {journal} {Nucl. Data Sheets},\ }\textbf
  {\bibinfo {volume} {111}},\ \bibinfo {pages} {2425 } (\bibinfo {year}
  {2010})}\BibitemShut {NoStop}%
\bibitem [{\citenamefont {Chadwick}\ \emph {et~al.}(2006)\citenamefont
  {Chadwick} \emph {et~al.}}]{endf}%
  \BibitemOpen
  \bibfield  {author} {\bibinfo {author} {\bibfnamefont {M.}~\bibnamefont
  {Chadwick}} \emph {et~al.},\ }\href
  {http://www.sciencedirect.com/science/article/pii/S0090375206000871}
  {\bibfield  {journal} {\bibinfo  {journal} {Nucl. Data Sheets},\ }\textbf
  {\bibinfo {volume} {107}},\ \bibinfo {pages} {2931 } (\bibinfo {year}
  {2006})},\ \bibinfo {note} {evaluated Nuclear Data File
  ENDF/B-VII.0}\BibitemShut {NoStop}%
\bibitem [{\citenamefont {Glazkov}(1963)}]{glazkov}%
  \BibitemOpen
  \bibfield  {author} {\bibinfo {author} {\bibfnamefont {N.~P.}\ \bibnamefont
  {Glazkov}},\ }\href@noop {} {\bibfield  {journal} {\bibinfo  {journal} {At.
  Energ. [Sov. J. At. Energy]},\ }\textbf {\bibinfo {volume} {15}},\ \bibinfo
  {pages} {416} (\bibinfo {year} {1963})},\ \bibinfo {note} {data taken from
  the EXFOR database, files EXFOR 40680.012 and 40680.013 dated Aug 15, 2007,
  retrieved from the IAEA Nuclear Data Services Website}\BibitemShut {NoStop}%
\bibitem [{\citenamefont {Hetrick}\ \emph {et~al.}(1984)\citenamefont
  {Hetrick}, \citenamefont {Fu},\ and\ \citenamefont {Larson}}]{cuendf}%
  \BibitemOpen
  \bibfield  {author} {\bibinfo {author} {\bibfnamefont {D.~M.}\ \bibnamefont
  {Hetrick}}, \bibinfo {author} {\bibfnamefont {C.~Y.}\ \bibnamefont {Fu}}, \
  and\ \bibinfo {author} {\bibfnamefont {D.~C.}\ \bibnamefont {Larson}},\
  }\href@noop {} {\emph {\bibinfo {title} {Calculated Neutron-Induced Cross
  Section for $^{63,65}${Cu} from 1 to 20 {MeV} and Comparisons with
  Experiments}}},\ \bibinfo {type} {Tech. Rep.}\ (\bibinfo  {institution} {Oak
  Ridge National Laboratory},\ \bibinfo {year} {1984})\ \bibinfo {note}
  {{ORNL/TM-9083, ENDF-337}}\BibitemShut {NoStop}%
\bibitem [{\citenamefont {Iwasaki}\ \emph {et~al.}(1979)\citenamefont
  {Iwasaki}, \citenamefont {Crawley}, \citenamefont {Markham}, \citenamefont
  {Finck},\ and\ \citenamefont {Kim}}]{iwa79}%
  \BibitemOpen
  \bibfield  {author} {\bibinfo {author} {\bibfnamefont {Y.}~\bibnamefont
  {Iwasaki}}, \bibinfo {author} {\bibfnamefont {G.~M.}\ \bibnamefont
  {Crawley}}, \bibinfo {author} {\bibfnamefont {R.~G.}\ \bibnamefont
  {Markham}}, \bibinfo {author} {\bibfnamefont {J.~E.}\ \bibnamefont {Finck}},
  \ and\ \bibinfo {author} {\bibfnamefont {J.~H.}\ \bibnamefont {Kim}},\ }\href
  {http://link.aps.org/doi/10.1103/PhysRevC.20.861} {\bibfield  {journal}
  {\bibinfo  {journal} {Phys. Rev. C},\ }\textbf {\bibinfo {volume} {20}},\
  \bibinfo {pages} {861} (\bibinfo {year} {1979})}\BibitemShut {NoStop}%
\bibitem [{\citenamefont {McCarthy}\ and\ \citenamefont
  {Crawley}(1966)}]{mcc66}%
  \BibitemOpen
  \bibfield  {author} {\bibinfo {author} {\bibfnamefont {A.~L.}\ \bibnamefont
  {McCarthy}}\ and\ \bibinfo {author} {\bibfnamefont {G.~M.}\ \bibnamefont
  {Crawley}},\ }\href {http://link.aps.org/doi/10.1103/PhysRev.150.935}
  {\bibfield  {journal} {\bibinfo  {journal} {Phys. Rev.},\ }\textbf {\bibinfo
  {volume} {150}},\ \bibinfo {pages} {935} (\bibinfo {year}
  {1966})}\BibitemShut {NoStop}%
\bibitem [{\citenamefont {Casten}(1990)}]{cast90}%
  \BibitemOpen
  \bibfield  {author} {\bibinfo {author} {\bibfnamefont {R.}~\bibnamefont
  {Casten}},\ }\href@noop {} {\emph {\bibinfo {title} {Nuclear structure from a
  simple perspective}}},\ Vol.~\bibinfo {volume} {13}\ (\bibinfo  {publisher}
  {Oxford University Press, USA},\ \bibinfo {year} {1990})\BibitemShut
  {NoStop}%
\bibitem [{\citenamefont {Agostinelli}\ \emph {et~al.}(2003)\citenamefont
  {Agostinelli} \emph {et~al.}}]{geant4}%
  \BibitemOpen
  \bibfield  {author} {\bibinfo {author} {\bibfnamefont {S.}~\bibnamefont
  {Agostinelli}} \emph {et~al.},\ }\href
  {http://www.sciencedirect.com/science/article/pii/S0168900203013688}
  {\bibfield  {journal} {\bibinfo  {journal} {Nucl. Instrum. Methods},\
  }\textbf {\bibinfo {volume} {A506}},\ \bibinfo {pages} {250 } (\bibinfo
  {year} {2003})}\BibitemShut {NoStop}%
\bibitem [{\citenamefont {Heikkinen}\ \emph {et~al.}(2008)\citenamefont
  {Heikkinen}, \citenamefont {Kaitaniemi},\ and\ \citenamefont
  {Boudard}}]{g4physics}%
  \BibitemOpen
  \bibfield  {author} {\bibinfo {author} {\bibfnamefont {A.}~\bibnamefont
  {Heikkinen}}, \bibinfo {author} {\bibfnamefont {P.}~\bibnamefont
  {Kaitaniemi}}, \ and\ \bibinfo {author} {\bibfnamefont {A.}~\bibnamefont
  {Boudard}},\ }\href {http://stacks.iop.org/1742-6596/119/i=3/a=032024}
  {\bibfield  {journal} {\bibinfo  {journal} {J. Phys. Conf. Proc.},\ }\textbf
  {\bibinfo {volume} {119}},\ \bibinfo {pages} {032024} (\bibinfo {year}
  {2008})}\BibitemShut {NoStop}%
\bibitem [{\citenamefont {Yoshida}\ \emph {et~al.}(2005)\citenamefont
  {Yoshida}, \citenamefont {Kishimoto}, \citenamefont {Ogawa}, \citenamefont
  {Hazama}, \citenamefont {Umehara}, \citenamefont {Matsuoka}, \citenamefont
  {Yokoyama}, \citenamefont {Mukaida}, \citenamefont {Ichihara},\ and\
  \citenamefont {Tatewaki}}]{candles}%
  \BibitemOpen
  \bibfield  {author} {\bibinfo {author} {\bibfnamefont {S.}~\bibnamefont
  {Yoshida}}, \bibinfo {author} {\bibfnamefont {T.}~\bibnamefont {Kishimoto}},
  \bibinfo {author} {\bibfnamefont {I.}~\bibnamefont {Ogawa}}, \bibinfo
  {author} {\bibfnamefont {R.}~\bibnamefont {Hazama}}, \bibinfo {author}
  {\bibfnamefont {S.}~\bibnamefont {Umehara}}, \bibinfo {author} {\bibfnamefont
  {K.}~\bibnamefont {Matsuoka}}, \bibinfo {author} {\bibfnamefont
  {D.}~\bibnamefont {Yokoyama}}, \bibinfo {author} {\bibfnamefont
  {K.}~\bibnamefont {Mukaida}}, \bibinfo {author} {\bibfnamefont
  {K.}~\bibnamefont {Ichihara}}, \ and\ \bibinfo {author} {\bibfnamefont
  {Y.}~\bibnamefont {Tatewaki}},\ }\href
  {http://www.sciencedirect.com/science/article/pii/S0920563204006164}
  {\bibfield  {journal} {\bibinfo  {journal} {Nucl. Phys. B, Proc. Suppl.},\
  }\textbf {\bibinfo {volume} {138}},\ \bibinfo {pages} {214 } (\bibinfo {year}
  {2005})},\ \bibinfo {note} {proceedings of the Eighth International Workshop
  on Topics in Astroparticle and Underground Physics}\BibitemShut {NoStop}%
\bibitem [{\citenamefont {Belogurov}\ \emph {et~al.}(2005)\citenamefont
  {Belogurov} \emph {et~al.}}]{2005Bel}%
  \BibitemOpen
  \bibfield  {author} {\bibinfo {author} {\bibfnamefont {S.}~\bibnamefont
  {Belogurov}} \emph {et~al.},\ }\href
  {http://dx.doi.org/10.1109/TNS.2005.852678} {\bibfield  {journal} {\bibinfo
  {journal} {IEEE Trans. Nucl. Sci.},\ }\textbf {\bibinfo {volume} {52}},\
  \bibinfo {pages} {1131 } (\bibinfo {year} {2005})}\BibitemShut {NoStop}%
\bibitem [{\citenamefont {Aalseth}\ \emph {et~al.}(2004)\citenamefont {Aalseth}
  \emph {et~al.}}]{2004Aal}%
  \BibitemOpen
  \bibfield  {author} {\bibinfo {author} {\bibfnamefont {C.~E.}\ \bibnamefont
  {Aalseth}} \emph {et~al.},\ }\href
  {http://www.springerlink.com/content/778x5202j2173631/} {\bibfield  {journal}
  {\bibinfo  {journal} {Phys. At. Nucl.},\ }\textbf {\bibinfo {volume} {67}},\
  \bibinfo {pages} {2002} (\bibinfo {year} {2004})}\BibitemShut {NoStop}%
\bibitem [{\citenamefont {{Kosyak}}\ \emph
  {et~al.}(2000){\natexlab{b}}\citenamefont {{Kosyak}}, \citenamefont
  {{Chekushina}},\ and\ \citenamefont {{Ermatov}}}]{2000KO}%
  \BibitemOpen
  \bibfield  {author} {\bibinfo {author} {\bibfnamefont {Y.~G.}\ \bibnamefont
  {{Kosyak}}}, \bibinfo {author} {\bibfnamefont {L.~V.}\ \bibnamefont
  {{Chekushina}}}, \ and\ \bibinfo {author} {\bibfnamefont {A.~S.}\
  \bibnamefont {{Ermatov}}},\ }in\ \href@noop {} {\emph {\bibinfo {booktitle}
  {Program and Thesis}}}\ (\bibinfo {year} {2000})\BibitemShut {NoStop}%
\bibitem [{\citenamefont {Nomachi}\ \emph {et~al.}(2005)\citenamefont {Nomachi}
  \emph {et~al.}}]{Moon}%
  \BibitemOpen
  \bibfield  {author} {\bibinfo {author} {\bibfnamefont {M.}~\bibnamefont
  {Nomachi}} \emph {et~al.},\ }\href
  {http://www.sciencedirect.com/science/article/pii/S0920563204006188}
  {\bibfield  {journal} {\bibinfo  {journal} {Nucl. Phys. B, Proc. Suppl.},\
  }\textbf {\bibinfo {volume} {138}},\ \bibinfo {pages} {221 } (\bibinfo {year}
  {2005})},\ \bibinfo {note} {proceedings of the Eighth International Workshop
  on Topics in Astroparticle and Underground Physics}\BibitemShut {NoStop}%
\bibitem [{\citenamefont {Arnold}\ \emph {et~al.}(2005)\citenamefont {Arnold}
  \emph {et~al.}}]{nemo}%
  \BibitemOpen
  \bibfield  {author} {\bibinfo {author} {\bibfnamefont {R.}~\bibnamefont
  {Arnold}} \emph {et~al.},\ }\href
  {http://www.sciencedirect.com/science/article/pii/S0168900204016821}
  {\bibfield  {journal} {\bibinfo  {journal} {Nucl. Instrum. Methods},\
  }\textbf {\bibinfo {volume} {A536}},\ \bibinfo {pages} {79 } (\bibinfo {year}
  {2005})}\BibitemShut {NoStop}%
\bibitem [{\citenamefont {Dolinski}\ \emph {et~al.}(2011)\citenamefont
  {Dolinski}, \citenamefont {Devlin}, \citenamefont {Fotiadis}, \citenamefont
  {Garrett}, \citenamefont {Nelson}, \citenamefont {Norman},\ and\
  \citenamefont {Younes}}]{dol11}%
  \BibitemOpen
  \bibfield  {author} {\bibinfo {author} {\bibfnamefont {M.}~\bibnamefont
  {Dolinski}}, \bibinfo {author} {\bibfnamefont {M.}~\bibnamefont {Devlin}},
  \bibinfo {author} {\bibfnamefont {N.}~\bibnamefont {Fotiadis}}, \bibinfo
  {author} {\bibfnamefont {P.}~\bibnamefont {Garrett}}, \bibinfo {author}
  {\bibfnamefont {R.}~\bibnamefont {Nelson}}, \bibinfo {author} {\bibfnamefont
  {E.}~\bibnamefont {Norman}}, \ and\ \bibinfo {author} {\bibfnamefont
  {W.}~\bibnamefont {Younes}},\ }\href
  {http://www.sciencedirect.com/science/article/pii/S0920563211007316}
  {\bibfield  {journal} {\bibinfo  {journal} {Nucl. Phys. B, Proc. Suppl.},\
  }\textbf {\bibinfo {volume} {221}},\ \bibinfo {pages} {341 } (\bibinfo {year}
  {2011})},\ \bibinfo {note} {the Proceedings of the 22nd International
  Conference on Neutrino Physics and Astrophysics}\BibitemShut {NoStop}%
\bibitem [{\citenamefont {Ackerman}\ \emph {et~al.}(2011)\citenamefont
  {Ackerman} \emph {et~al.}}]{exo2bb}%
  \BibitemOpen
  \bibfield  {author} {\bibinfo {author} {\bibfnamefont {N.}~\bibnamefont
  {Ackerman}} \emph {et~al.} (\bibinfo {collaboration} {EXO Collaboration}),\
  }\href {http://link.aps.org/doi/10.1103/PhysRevLett.107.212501} {\bibfield
  {journal} {\bibinfo  {journal} {Phys. Rev. Lett.},\ }\textbf {\bibinfo
  {volume} {107}},\ \bibinfo {pages} {212501} (\bibinfo {year}
  {2011})}\BibitemShut {NoStop}%
\bibitem [{\citenamefont {Chen}(2007)}]{snoplus}%
  \BibitemOpen
  \bibfield  {author} {\bibinfo {author} {\bibfnamefont {M.~C.}\ \bibnamefont
  {Chen}},\ }\href {http://link.aip.org/link/?APC/944/25/1} {\bibfield
  {journal} {\bibinfo  {journal} {AIP Conf. Proc.},\ }\textbf {\bibinfo
  {volume} {944}},\ \bibinfo {pages} {25} (\bibinfo {year} {2007})}\BibitemShut
  {NoStop}%
\end{thebibliography}%

\end{document}